\documentclass[acmlarge, screen]{acmart}

\usepackage{titlesec}
\usepackage{balance}
\usepackage[commandnameprefix=always]{changes}
\usepackage{booktabs}
\usepackage{multirow}
\usepackage{xcolor}

\usepackage{float}
\usepackage{algorithm}
\usepackage{algpseudocode}
\usepackage{hyperref}
\usepackage{url}
\usepackage{algorithm}
\usepackage{graphicx}
\usepackage{subcaption}
\usepackage{algpseudocode}
\usepackage{amsmath}
\usepackage{geometry}

\usepackage{etoolbox}  
\usepackage{tikz}      
\usepackage{lipsum}
\usepackage{enumitem}
\usepackage{tabularx}
\usepackage{array}

\AtBeginDocument{%
  }

\setcopyright{cc}
\setcctype{by-nc-nd}
\acmJournal{IMWUT}
\acmYear{2026} \acmVolume{10} \acmNumber{3} \acmArticle{80}
\acmMonth{9} \acmDOI{10.1145/3832043}

\begin{document}
\title{SocialPulse: On-Device Detection of Social Interactions in Naturalistic Settings Using Smartwatch Sensing}
 
\author{Md Sabbir Ahmed}
\authornote{Corresponding author}
\email{msabbir@virginia.edu}
\affiliation{%
  \institution{Department of Systems and Information Engineering, University of Virginia}
  \country{USA}
}

\author{Kaitlyn Dorothy Petz}
\email{kdp8y@virginia.edu}
\affiliation{%
  \institution{Department of Psychology, University of Virginia}
  \country{USA}
}

\author{Noah French}
\email{njf5cu@virginia.edu}
\affiliation{%
  \institution{Department of Psychology, University of Virginia}
  \country{USA}
}

\author{Tanvi Lakhtakia}
\email{lakhtakia@virginia.edu}
\affiliation{%
  \institution{Department of Psychology, University of Virginia}
  \country{USA}
}

\author{Aayushi Sangani}
\email{cjv8xy@virginia.edu}
\affiliation{%
  \institution{Department of Psychology, University of Virginia}
  \country{USA}
}

\author{Mark Rucker}
\email{mr2an@virginia.edu}
\affiliation{%
  \institution{Department of Systems and Information Engineering, University of Virginia}
  \country{USA}
}

\author{Xinyu Chen}
\email{dfs3mc@virginia.edu}
\affiliation{%
  \institution{Department of Systems and Information Engineering, University of Virginia}
  \country{USA}
}

\author{Bethany A. Teachman}
\email{bteachman@virginia.edu}
\affiliation{%
  \institution{Department of Psychology, University of Virginia}
  \country{USA}
}

\author{Laura E. Barnes}
\email{lb3dp@virginia.edu}
\affiliation{%
  \institution{Department of Systems and Information Engineering, University of Virginia}
  \country{USA}
}

\begin{abstract}

Social interactions are fundamental to well-being, yet automatically detecting them in daily life—particularly using wearables—remains underexplored. Most existing systems are evaluated in controlled settings, focus primarily on in-person interactions, or rely on restrictive assumptions (e.g., requiring multiple speakers within fixed temporal windows), limiting generalizability to real-world use. We present an on-watch interaction detection system designed to capture diverse interactions in naturalistic settings. A core component is a foreground speech detector trained on a public dataset. Evaluated on over 100{,}000 labeled foreground speech and background sound instances, the detector achieves a balanced accuracy of 85.51\%, outperforming prior work by 5.11\%.

We evaluated the system in a real-world deployment (N=38), with over 900 hours of total smartwatch wear time. The system detected 1{,}691 interactions, 77.28\% were confirmed via participant self-report, with durations ranging from under one minute to over one hour. Among correct detections, 81.45\% were in-person, 15.7\% virtual, and 1.85\% hybrid. We further developed a 15-second window-level audio-only model that enables faster interaction prediction, achieving a balanced accuracy of 90.39\% and a sensitivity of 91.01\% on 33{,}698 labeled windows. These results demonstrate the feasibility of real-world interaction sensing and open the door to adaptive, context-aware systems responding to users’ dynamic social environments.

\end{abstract}



\maketitle

\section{Introduction}

Social interactions are an integral part of everyday life and play a key role in well-being \cite{Liang2024-nt, ono2011relationship}. Positive interactions with others can boost mood and create a sense of meaning \cite{park2015group}. However, negative social interactions, or complete social avoidance, can adversely affect health \cite{Shahrestani2015-dm, Schwerdtfeger2009-jy}, contributing to anxiety and depression \cite{Gooding2022, RASSABY2024102931}. Being able to detect real-world social interactions may allow us to disrupt unhealthy interaction patterns as they occur, enabling personalized interventions that promote mental well-being in everyday life. More broadly, social interaction detection is a foundational capability for context-aware systems, allowing them to respond dynamically to users' changing social contexts.

One method for capturing information about social interactions and their patterns is through self-report data (e.g., \cite{sun2020well, lopes2004emotional}). For example, an individual could be prompted periodically or at the end of the day to describe all of their social interactions. While straightforward to collect using traditional survey methods, such data are impacted by retrospective bias and may be burdensome for participants to provide \cite{Lucas2021-bk, Roos2023-wk}. A promising alternative is passive sensing, which can collect real-time data unobtrusively and with minimal user burden. Several studies have used smartphones to passively detect social interactions, but these are limited by small sample sizes (e.g., N=5 in \cite{Lane_2011}) and evaluations in controlled environments \cite{Katevas_Hänsel_Clegg_Leontiadis_Haddadi_Tokarchuk_2019}, raising concerns about generalizability. Further, smartphones may not collect high quality data important to interaction detection. For example, they are often stored in pockets, resulting in lower quality acoustic data \cite{Liang2023-gi}. In contrast, smartwatches are more consistently worn close to the body in different contexts and locations \cite{Shahmohammadi2017-hm}, which may make them more suitable for continuous interaction detection.

Despite growing interest in smartwatch-based social interaction detection, relatively few systems have been demonstrated and evaluated for real-time use in unconstrained, everyday settings, and existing approaches necessarily make design choices that shape their scope and applicability. For example, Liang et al.~\cite{Liang2023-gi} present foundational work on a smartwatch system for detecting in-person social interactions, demonstrating the promise of on-watch sensing for social context inference. Their system focuses specifically on in-person interactions and adopts modeling assumptions —such as requiring the watch user with at least one other person within fixed temporal windows— that are well suited to certain conversational settings but may not generalize to others, like interactions with extended single-speaker segments, as well as virtual (e.g., phone or video calls) and hybrid interactions, in which participants engage with a combination of co-located and remote conversation partners, which have become an increasingly common part of daily life \cite{Hutchins_Allen_Curran_Kannis-Dymand_2021}.


Other work has explored alternative sensing modalities, including physiological signals such as respiration, to detect conversational activity \cite{Rahman2011-kx, Bari2018-os}. While these approaches provide valuable insights into the relationship between physiology and social interaction, they face practical challenges for large-scale, real-world deployment. In particular, respiration sensors are not available on most commercially deployed smartwatches (e.g., Samsung Galaxy Watch~\cite{Wikipedia_contributors2025-qs}), and respiration rate estimates on devices such as Fitbit and Apple Watch are typically limited to sleep periods \cite{noauthor-fitbit-breathing-rate, noauthor-undated-av-apple}, when social interactions are unlikely to occur.

To build on foundational smartwatch-based systems and address their limitations, we propose a social interaction detection system that operates on off-the-shelf Wear OS smartwatches and captures audio-based interactions across in-person, virtual, and hybrid settings. In addition to enabling real-time interaction detection, the system was explicitly designed to facilitate the collection of fine-grained, real-world interaction labels. To this end, we developed an initial prototype for automatic interaction detection (Section~\ref{initial_system}).

The system supports participant-driven annotation by leveraging smartwatch-based ecological momentary assessments (EMAs) that are triggered when potential interaction start and end times are detected based on probability of foreground speech (section \ref{algo_interaction_section}). This design directly addresses a key challenge in naturalistic data collection: users may forget or be unwilling to manually initiate or terminate recordings during everyday interactions. By prompting users at salient moments, the smartwatch serves both as a sensing platform and as a low-burden labeling interface. We deployed the system on smartwatches (Section~\ref{on_watch_imple_section}) with 38 participants (Section~\ref{real_world_deploy_section}) to evaluate detection performance while simultaneously collecting annotated interaction labels alongside multimodal sensor data (Section~\ref{sensing_config}, e.g., photoplethysmography). To further improve label coverage, we employed multiple complementary annotation strategies, including periodic prompts to identify missed interactions (Section~\ref{all_annot_for_interact_section}).

Results from the on-device deployment of our initial prototype (Section~\ref{real_world_deploy_section}) show that participants used the system for a total of 120 days. During this period, the system automatically detected 1{,}691 interactions, achieving an interaction detection accuracy of 77.28\% ($N = 1{,}299$), with an average per-participant accuracy of 79.56\%. The correctly detected interactions spanned in-person, virtual, and hybrid settings and covered a wide range of durations, from brief interactions (approximately one minute) to extended interactions lasting over one hour, demonstrating the system’s ability to capture diverse social interactions in fully naturalistic settings. 

A limitation of the deployed framework is that it confirms an interaction only after the possible interaction interval ends, which can delay detection for applications such as adaptive sensing and context-aware systems. To support faster detection, we trained 15-second window-level models (Section \ref{multi_modal_section}) using labels from the real-world deployment, enabling interaction prediction from only 15 seconds of data. This was not feasible before deployment, as only semi-controlled foreground-speech labels were available. The multimodal model with the meta-learner achieved a balanced accuracy of 90.36\% and sensitivity of 91.17\% on 33,698 labeled 15-second windows collected in fully naturalistic settings (Section \ref{perf_multi_mod_section}). Interestingly, the audio-only meta-learner, trained on the same 15-second windowed data and relying solely on audio-derived inputs, achieved similar performance with a balanced accuracy of 90.39\% and sensitivity of 91.01\% (Section \ref{audio_perf_on_15_sec_windows}). These findings indicate that audio is the dominant modality and that the additional sensors do not substantially improve prediction performance in this setting. The audio-only model demonstrates that robust, low-latency detection can be achieved using a minimal audio-only sensing configuration.

Through this work, we contribute to the ubiquitous computing and human--computer interaction (HCI) communities in the following ways:
\begin{itemize}[noitemsep, topsep=0pt]
\item \textbf{Improved foreground speech detection for on-device social sensing.}  
We introduce a foreground speech detector (FSD) that achieves a balanced accuracy of 85.51\%, outperforming recent state-of-the-art approaches~\cite{Liang_Xu_Chen_Adaimi_Harwath_Thomaz_2023, Ahmed_Rahman_Wang_Rucker_Barnes_2024} by approximately 5.11\%. As foreground speech detection is a core building block for privacy-preserving social sensing tasks—including social interaction detection, social connection analysis, and network estimation—this improvement has broad applicability within ubiquitous computing systems.
\item \textbf{A fully on-device, real-time system for social interaction detection in daily life.} We present and validate an end-to-end smartwatch-based system that detects social interactions in real-time using fully on-device inference and privacy-preserving sensing. Raw audio is processed transiently on-device and is neither stored permanently nor transmitted. We evaluate the system in a three-day real-world deployment with 38 participants, demonstrating reliable detection of \emph{in-person, virtual, and hybrid} interactions ranging from approximately one minute to over one hour in duration. Compared to prior on-device interaction detection systems~\cite{Liang2023-gi, Zhang_Bertley_Liang_Thomaz_2025}, our system advances the state of the art in several important ways:
    \begin{itemize}
        \item It operates without restrictive assumptions, such as requiring the watch user with at least one other person within fixed temporal windows.
        \item To our knowledge, it represents the \emph{largest real-world evaluation} of an on-watch interaction detection system to date, comprising over 900 hours of continuous smartwatch operation, compared to fewer than 50 hours in prior work~\cite{Liang2023-gi}.
        \item It supports detection of in-person, virtual, and hybrid interactions, extending prior smartwatch-based approaches that primarily focus on in-person encounters~\cite{Liang2023-gi, Zhang_Bertley_Liang_Thomaz_2025}.
    \end{itemize}
\item \textbf{Resource efficient interaction detection system with strong accuracy--efficiency tradeoffs.} Trained on real-world collected data, we further developed a 15-second window-level interaction detection model that supports faster identification for applications such as adaptive sensing and context-aware systems. For that purpose, we introduce an audio-only interaction detection model that achieves a balanced accuracy of 90.39\% when evaluated on more than 900 hours of fully naturalistic data containing 33,698 samples. This represents improvements of \emph{9.26\%} and \emph{2.92\%} in balanced accuracy compared to models trained using MobileNetV1~\cite{mobile_nets} and a custom YAMNet-based~\cite{YAMNet} model, respectively. Importantly, the proposed system design is resource efficient and reduces resource consumption relative to existing approaches, enabling continuous, real-time interaction detection on smartwatches \emph{without requiring raw audio transmission, storage, or off-device processing}.
\item \textbf{A fully naturalistic smartwatch dataset for real-world social interaction analysis.}
We open-source our dataset\footnote{https://zenodo.org/records/19767766}, which includes audio features for each data collection window, corresponding interaction labels, and anonymized participant IDs to support fine-grained analysis. To the best of our knowledge, this is the first publicly available dataset containing wearable sensing features and labels for social interactions captured entirely in naturalistic daily-life environments. In doing so, it extends prior work~\cite{Liang2023-gi, Zhang_Bertley_Liang_Thomaz_2025}, which were either collected in semi-naturalistic settings or span fewer than 50 hours.

\end{itemize}

\section{Related Work}

Prior work on social interaction detection spans smartphone- and wearable-based systems. Many existing approaches, however, rely on continuous audio capturing, off-device processing, or restrictive assumptions that limit privacy and real-world generalizability. Motivated by mental health applications requiring reliable, in-the-wild sensing under strict privacy and resource constraints, we organize related work by platform (smartphones vs. wearables) and sensing modality, emphasizing on-device feasibility, privacy, and modeling assumptions.


\subsection{Interaction Detection Leveraging Smartphones}

Early work on social interaction detection relied on custom-built devices prior to the widespread adoption of smartphones \cite{choudhury2004sociometer}. The emergence of smartphones enabled broader exploration of interaction sensing using embedded sensors, particularly audio. Several studies leveraged speech and voice activity detection to infer conversations and social interactions \cite{Rabbi_Ali_Choudhury_Berke_2011, Lane_2011, Roos2023-wk}, with later systems integrating multimodal sensing \cite{Katevas_Hänsel_Clegg_Leontiadis_Haddadi_Tokarchuk_2019}. Models proposed in early work \cite{Rabbi_Ali_Choudhury_Berke_2011, Lane_2011} informed the conversation detection module in the StudentLife project \cite{Wang_Chen_Chen_Li_Harari_Tignor_Zhou_Ben-Zeev_Campbell_2014}, which subsequently formed the basis of the AWARE framework \cite{noauthor_undated-aware_framework} and has been widely adopted in phone-based studies \cite{Roos2023-wk}.

Despite their influence, most smartphone-based approaches were evaluated using limited labeled data collected under controlled conditions \cite{ Luo_Chan_2013, Lu2011-wg, Ahmed2015-bn} or very short naturalistic settings \cite{Luo_Chan_2013}, leaving their robustness in naturalistic settings unclear. In addition, these systems rely primarily on speech related features, which can reduce reliability in real-world contexts where speech is intermittent or absent. While non-speech vocalizations or paralinguistic cues such as laughter or silence may also indicate social interaction, they are typically not explicitly modeled. Taken together, smartphones, while pervasive, are not consistently on-person or uniquely associated with a single user and are often obstructed, whereas smartwatches are worn more continuously, can capture more interactions, and support higher notification response rates, making them better suited for reliable social sensing and low-burden interventions \cite{Shahmohammadi2017-hm, King_Sarrafzadeh_2017, Liang2023-gi, Ponnada_Haynes_Maniar_Manjourides_Intille_2017}.

\subsection{Interaction Detection Leveraging Wearable Sensors}
\label{related_work_wear}
Prior work has explored wearable sensing for detecting conversational activity, including respiration-based bands \cite{Rahman2011-kx, Bari2018-os}. Several studies collected data in both controlled and real-world environments \cite{Bari2018-os, Bari2020-jd}. Because labeling in-the-wild data is challenging, some work used speech segments detected by the LENA device as ground truth for interaction boundaries \cite{Bari2018-os}, later combining LENA with learned models \cite{Bari2020-jd}. However, LENA was originally designed for child speech and relies on simple Gaussian mixture models \cite{Xu2008SignalPF, Richards2010LENA}, which limits its performance on adult populations. Recent evaluations report low recall (e.g., 62\% \cite{Meera_Murali_Raju_Srikar_Shyam_Rao_et_al._2025}) for adult speaker recognition \cite{Meera_Murali_Raju_Srikar_Shyam_Rao_et_al._2025, Lehet_Arjmandi_Houston_Dilley_2020}. Moreover, respiration-based systems typically require additional hardware such as chest straps \cite{Rahman2011-kx, Bari2018-os}, which can be obtrusive and impractical for long-term, everyday use. These systems also lack scalability, as off-the-shelf smartwatches do not include respiration sensors, and commercial respiration estimates are generally limited to sleep contexts \cite{noauthor-fitbit-breathing-rate, noauthor-undated-av-apple}.

To address these limitations, more recent work has investigated social sensing using commercially available smartwatches. For example, a recent study \cite{Dhand_Tate_SocialBit_Dahima_etal2026} introduced the SocialBit algorithm for interaction detection. While the approach demonstrated strong clinical validation, its generalizability to naturalistic home and community settings remains uncertain because the data were collected from stroke patients in a hospital environment. In addition, reliance on continuous livestream video observation may reduce ecological validity, as participants may alter their behavior when being observed. Finally, sampling audio every five minutes increases the likelihood of missing brief or transient social interactions. \cite{Boateng2019-gt} developed an on-watch speech detection system using 3.5 hours of data. However, speech alone is insufficient for reliable interaction detection: a system needs to distinguish foreground speech from background or ambient audio \cite{Liang2023-gi}. Without this distinction, environmental sounds such as television audio or nearby conversations may be misclassified as social interactions \cite{Lane_2011}. Consequently, foreground speech detection has emerged as a critical component of social interaction sensing.

Several models for foreground speech detection have been proposed \cite{Ahmed_Rahman_Wang_Rucker_Barnes_2024, Liang_Xu_Chen_Adaimi_Harwath_Thomaz_2023, Nadarajan_Somandepalli_Narayanan_2019}. However, some foreground speech–based interaction detection approaches impose assumptions that may not hold across all real-world contexts. For example, the system described in~\cite{Liang2023-gi} assumes the presence of at least two speakers, including the watch user, within fixed temporal windows, potentially limiting performance in scenarios such as one-on-one meetings or lectures.

Beyond audio, prior work has explored additional sensing modalities for interaction inference. Physiological signals have been shown to differ between interaction and non-interaction contexts \cite{Hänsel_Katevas_Orgs_Richardson_Alomainy_Haddadi_2018}, and IMU data may provide complementary behavioral cues. A recent study combined acoustic and IMU sensing for social interaction detection \cite{Zhang_Bertley_Liang_Thomaz_2025}, but similarly assumed presence of watch user with at least one other speaker within short windows and evaluated performance in semi-controlled, speech-centric settings. As a result, the generalizability of these approaches to unconstrained, naturalistic environments remains an open challenge.

\section{Design of Social Interaction Detection Pipeline for In-the-Wild Annotation}
\label{initial_system}

Before collecting new real-world participant data, we developed and deployed an initial social interaction detection pipeline on off-the-shelf smartwatches using a publicly available dataset and a pretrained model. This pipeline ran on-device during participants’ daily lives and was used to support in-the-wild annotation. When the system detected the end of a potential social interaction, it prompted users to confirm whether an interaction had occurred, enabling fine-grained, event-level annotation with minimal burden. To complement these event-triggered prompts, the system also issued periodic notifications (every 80–90 minutes), asking participants whether any interactions had been missed during the preceding interval and, if so, to report their approximate start and end times. This combination of real-time, on-device detection with event-driven and periodic prompting was designed to balance annotation burden with coverage, improving the capture of naturally occurring social interactions in daily life.

\subsection{Defining and Operationalizing Social Interaction}
\label{define_interact}

We define social interaction as vocal engagement with others occurring in person or through virtual modalities (e.g., phone or video calls), consistent with prior work \cite{Zhaoyang_Sliwinski_Martire_Smyth_2018}. Unlike previous approaches that impose additional constraints such as stationary participants or continuous turn-taking speech \cite{Katevas_Hänsel_Clegg_Leontiadis_Haddadi_Tokarchuk_2019}, we do not require such restrictions to better capture the variety of real-world interactions that can occur.

Although the definition of social interaction may overlap with conversation, where conversation is typically defined as “verbal interaction between two or more people” \cite{Yeomans_Boland_Collins_Abi-Esber_Brooks_2023}, our system is designed to detect social interaction broadly rather than conversation alone. Prior systems, such as \cite{Liang2023-gi, Zhang_Bertley_Liang_Thomaz_2025}, focus on conversation detection using continuous sensing and require the presence from at least two speakers, including the watch wearer, within every 30-second window. We build on this foundational work, as such assumptions may not consistently hold in real-world social interactions (e.g., during student–advisor meetings, storytelling where one individual speaks for extended periods, or situations where two friends talk while the watch user mainly listens and speaks later). 

Crucially, social interaction is not characterized by continuous speech alone. Paralinguistic vocal behaviors (e.g., laughter, shouting) and periods of silence may still reflect reciprocal engagement between individuals. Accordingly, instead of imposing a per-window requirement that the watch wearer speak during every 15-second sensing window, we require a minimum amount of foreground speech from the wearer aggregated over the duration of an interaction (e.g., across a 20-minute interval), better reflecting real-world interaction dynamics in which individual airtime varies substantially \cite{Cooney_Mastroianni_Abi-Esber_Brooks_2020}. Moreover, the system incorporates cues such as laughter, shouting, and whispering, as well as periods of silence.

The system also employs duty-cycled sensing rather than continuous recording, collecting data for 15 seconds every 1 minute and 30 seconds. This design allows interaction-related behaviors, both verbal and non-verbal, to occur during unsensed intervals, further aligning detection with naturalistic patterns of social interaction. 

Taken together, these choices move beyond conversation-centric assumptions and align with broader conceptualizations of social interaction \cite{Zhaoyang_Sliwinski_Martire_Smyth_2018}, introducing substantial flexibility and enabling the system to capture situations in which individuals are socially engaged without continuously conversing. 

\subsection{Development and Evaluation of Foreground Speech Detection}
\label{fsd_section}

\subsubsection{Model Development.} 
\label{fsd_model_dev_section}
Consistent with our definition of social interaction \cite{Zhaoyang_Sliwinski_Martire_Smyth_2018}, the presence of the watch wearer’s own speech constitutes a necessary signal for interaction detection. Accordingly, we developed an FSD that identifies speech produced by the watch wearer while treating all other audio, including speech from others, music, environmental sounds, and silence, as background \cite{Ahmed_Rahman_Wang_Rucker_Barnes_2024, Liang_Xu_Chen_Adaimi_Harwath_Thomaz_2023}.

\textit{Dataset and Constraints.} 
We trained the model using a publicly available dataset \cite{Liang_Xu_Chen_Adaimi_Harwath_Thomaz_2023} consisting of 1-second audio segments, each manually labeled as either foreground or background sound. The dataset contains a total of 113,820 instances, of which 76.29\% ($N=86{,}829$) are labeled as background sound and 23.71\% ($N=26{,}991$) as foreground speech.

One approach to foreground speech detection relies on hand-crafted acoustic features (e.g., MFCCs and chromagrams) \cite{Ahmed_Rahman_Wang_Rucker_Barnes_2024}. While extracting multiple features on-device can support richer behavioral characterization and enable reuse across downstream tasks, such feature extraction can incur substantial computational overhead. For example, prior work reports that feature extraction is required over 1.5 minutes for nearly every probe start \cite{Ahmed_Rahman_Wang_Rucker_Barnes_2024}, limiting its suitability for scenarios in which timely or continuous prediction is required.

Although classical machine learning models (e.g., decision trees) may train more efficiently than deep learning models, the feature-extraction stage remains computationally expensive and can also constrain predictive performance. Moreover, our objective is not to train models on-device. Instead, we train models offline and deploy them to smartwatches for real-time inference, explicitly minimizing on-device computation to support sustained, long-term operation on resource-constrained wearable devices.

In developing the deep learning model for FSD, we focused on leveraging transfer learning, as prior work has demonstrated its effectiveness in wearable and audio-based sensing tasks \cite{Ebbehoj_Thunbo_Andersen_Glindtvad_Hulman_2022, WatchAnxiety_Ahmed_Barnes_2025} and suggests it may similarly benefit foreground speech identification.

\textit{Embedding-Based Modeling with Transfer Learning}: To develop the FSD, we extracted embeddings from the YAMNet model \cite{YAMNet}. YAMNet is a suitable choice because it is trained on 521 audio event classes (e.g., speech, laughter, music) from the large-scale AudioSet dataset \cite{audio_set_dataset, audioset_website}, which comprises 2.1 million YouTube videos totaling 5.8 thousand hours of audio. Training on such a large and diverse set of sound classes enables YAMNet to learn general-purpose audio representations. In addition, YAMNet is computationally efficient, with reported inference latency below 15 milliseconds (ms) \cite{yamnet_latency}.

\begin{figure}
    \centering
    \includegraphics[width=1\linewidth]{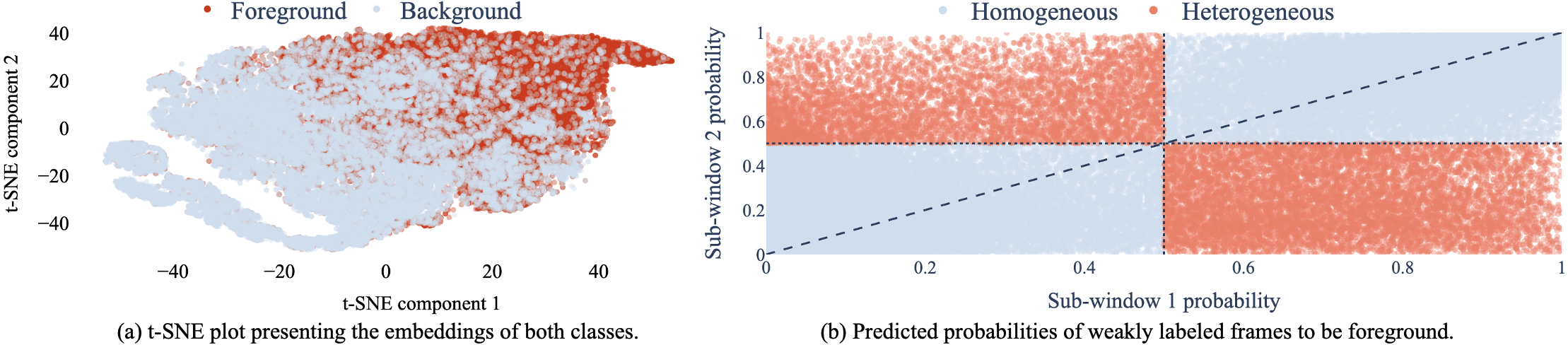}
    \caption{Difference between foreground and background sound. 
(a) Visualization of embeddings for 50{,}000 audio frames (25{,}000 per class). For readability, embeddings for all 227{,}640 frames are not shown. 
(b) Predicted foreground speech probabilities for all frames. Homogeneous blocks indicate audio instances in which both sub-frames have predicted probabilities either above or below 50\%, whereas heterogeneous blocks indicate instances in which the two sub-frames have opposite predictions (e.g., one sub-frame exceeds the 50\% threshold while the other falls below it).}
    \label{fig:fsd_embeddings_prob}
\end{figure}

By visualizing the embeddings of the dataset used to train the FSD model, we observed that YAMNet embeddings exhibit class-discriminative structure between foreground and background audio segments, but with substantial overlap (Figure~\ref{fig:fsd_embeddings_prob}(a)). This overlap suggests that simple linear decision boundaries are insufficient, motivating the use of a neural network model capable of learning non-linear relationships.

Accordingly, we developed a fully connected neural network classifier operating on the extracted embeddings to predict whether an audio instance corresponds to foreground speech or background sound. We used binary cross-entropy as the loss function and the Adam optimizer for training. To address class imbalance and ensure that the minority class received sufficient emphasis during learning, we applied class weighting using the formula
$
\text{weight}_c = \frac{1}{n_c} \cdot \frac{N}{2},
$
where $n_c$ denotes the number of instances in class $c$ and $N$ is the total number of instances. This weighting strategy balances the loss contributions across classes and mitigates bias toward the majority class.

The model was trained for 25 epochs, with early stopping applied when the validation loss did not improve for three consecutive epochs.

\textit{Weak Supervision and Meta-Learning.} 
Each 1-second segment in the dataset used to train the FSD model corresponds to two YAMNet frames, as YAMNet processes fixed 0.48-second windows regardless of input length. Although it may seem reasonable to train and evaluate models directly on these 0.48-second subwindows by assigning them the same label as the parent 1-second segment, this approach can be misleading. As noted in the dataset description \cite{Liang_Xu_Chen_Adaimi_Harwath_Thomaz_2023}, a 1-second segment is labeled as foreground speech if it contains either only the watch wearer’s speech or a mixture of the wearer’s and others’ speech. Consequently, individual subwindows within a mixed segment may not correspond to purely foreground speech. Assigning identical labels to such subwindows and using them for evaluation can therefore introduce label noise and distort performance estimates.

To maintain consistency with the dataset’s instance-level labeling scheme while leveraging pretrained embeddings, we adopted a weak supervision strategy. Specifically, we assigned the 1-second label to each of the two subwindows and trained the deep learning model on these weakly labeled subwindows to obtain predicted probabilities for each frame. We then trained a meta-learner that used the two subwindow probability outputs as input features to produce a final classification for the full 1-second audio instance (foreground vs.\ background), which was used for model evaluation.

A straightforward alternative would be to determine the final class based solely on the direction of the subwindow predictions; for example, labeling an instance as foreground speech only if both subwindow probabilities exceed 0.5. However, as shown in Figure~\ref{fig:fsd_embeddings_prob}(b), many instances exhibit disagreement between the two subwindows. In such cases, a naive rule based only on prediction direction would fail to classify a substantial number of instances accurately. Accordingly, we evaluated multiple classification algorithms for the meta-learner, using the predicted subwindow probabilities as input features.

\begin{figure}
    \centering
\includegraphics[width=1\linewidth]{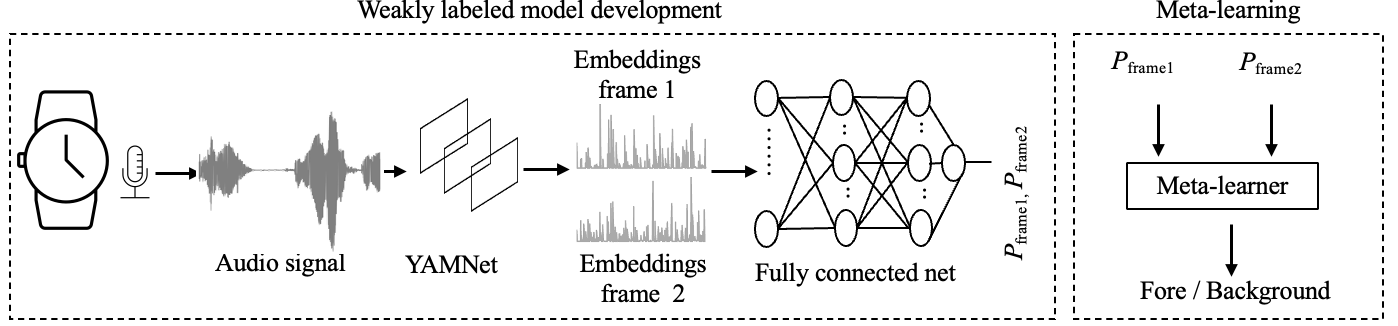}
    \caption {The pipeline for on-watch foreground speech prediction. \(P_{frame1}\) and  \(P_{frame2}\) denote the probabilities of foreground speech for frame 1 and frame 2, respectively.}
    \label{fig:fsd_architec}
\end{figure}

\subsubsection{Model Evaluation and Baselines.}

For evaluation, we employed a stratified 10-fold cross-validation procedure. In each iteration, nine folds were used for training, while the remaining fold was split evenly: 50\% of its samples were used for validation and 50\% for testing. We repeated this process by swapping the validation and test subsets so that, across iterations, the entire fold was treated as test data. Importantly, the model was retrained from scratch for each split to prevent information leakage between the validation and test sets. To maximize the amount of training data, no validation samples were drawn from the nine training folds. For the meta-learning stage, we maintained the same training and testing folds as those used for the weakly labeled model to ensure consistency and to avoid information leakage across learning stages.

To benchmark our foreground speech detection models, we compared them against two existing approaches \cite{Liang_Xu_Chen_Adaimi_Harwath_Thomaz_2023, Ahmed_Rahman_Wang_Rucker_Barnes_2024}, both of which were evaluated on the same publicly available dataset \cite{Liang_Xu_Chen_Adaimi_Harwath_Thomaz_2023}. The method proposed by \cite{Liang_Xu_Chen_Adaimi_Harwath_Thomaz_2023} employs a convolutional neural network–based architecture, whereas \cite{Ahmed_Rahman_Wang_Rucker_Barnes_2024} uses classical machine learning models. 

We evaluated a diverse set of classification algorithms, including Decision Trees, K-Nearest Neighbors (KNN), and Logistic Regression, as candidates for the meta-learner trained on predicted probabilities from 113{,}820 audio instances labeled as foreground or background speech. Among these approaches, a Nearest Centroid–based classifier \cite{nearest_centroid} achieved the best performance, attaining a balanced accuracy of 85.51\%. By comparison, the existing comparison models evaluated on the same dataset report balanced accuracies of 80.4\% \cite{Liang_Xu_Chen_Adaimi_Harwath_Thomaz_2023} and 80.3\% \cite{Ahmed_Rahman_Wang_Rucker_Barnes_2024}; our approach therefore improves balanced accuracy by at least 5.11\%. 

\begin{table}[!htbp]
\centering
\begin{tabular}{lccccc}
\toprule
Algorithm & Precision & Sensitivity & Specificity & BA & F1 \\
\midrule
    Meta-learner (DecisionTree) & 75.98 & 63.50 & 88.56 & 76.03 & 76.00 \\
    Meta-learner (ExtraTrees) & 80.42 & 66.94 & 91.47 & 79.21 & 79.78 \\
    Meta-learner (RandomForest) & 81.04 & 67.58 & 91.82 & 79.70 & 80.33 \\
    Meta-learner (KNN) & 81.04 & 68.23 & 91.67 & 79.95 & 80.47 \\
    Meta-learner (GradientBoost) & 83.26 & 71.35 & 92.72 & 82.03 & 82.61 \\
    Meta-learner (MLP) & 83.24 & 71.67 & 92.64 & 82.15 & 82.67 \\
    Meta-learner (Logit) & 82.79 & 73.01 & 92.00 & 82.50 & 82.65 \\
    Meta-learner (LinearSVC) & 82.36 & 75.75 & 91.01 & 83.38 & 82.85 \\
    \textbf{Meta-learner (NearestCentroid)} & \textbf{80.21} & \textbf{84.86} & \textbf{86.16} & \textbf{85.51} & \textbf{82.14} \\
    \bottomrule
    Ahmed et al. 2024 \cite{Ahmed_Rahman_Wang_Rucker_Barnes_2024} & 73.0 & 85.3 & 75.2 & 80.3 & 74.0 \\
    Liang et al., 2023 \cite{Liang_Xu_Chen_Adaimi_Harwath_Thomaz_2023} & NA & NA & NA & 80.4 & 81.50 \\
\bottomrule
\end{tabular}
\caption{Performance comparison of meta-learning algorithms used in the proposed framework and existing baseline models \cite{Ahmed_Rahman_Wang_Rucker_Barnes_2024, Liang_Xu_Chen_Adaimi_Harwath_Thomaz_2023}. Values in bold indicate the best-performing model in terms of balanced accuracy. BA: Balanced Accuracy; NA: Not Available.}
\label{fsd_performance}
\end{table}

\subsubsection{Foreground Speech Detection Discussion.}

The improved performance of our foreground speech detection pipeline stems from two key design choices. First, we leveraged YAMNet embeddings trained on thousands of hours of diverse audio data, providing a more robust and generalizable representation than prior approaches based on manually engineered features or datasets collected in constrained contexts \cite{Ahmed_Rahman_Wang_Rucker_Barnes_2024, Liang_Xu_Chen_Adaimi_Harwath_Thomaz_2023, audio_set_dataset}. Second, we introduced a meta-learning stage that jointly reasons over predictions from the two subwindows within each 1-second segment. Rather than relying on a fixed threshold or naive aggregation rule, the meta-learner learns consistent patterns across frames, leading to more reliable foreground speech detection and a balanced accuracy of 85.51\%.

However, a limitation of our approach is that \cite{Liang_Xu_Chen_Adaimi_Harwath_Thomaz_2023} employed a leave-one-group-out validation scheme, whereas we used a ten-fold cross-validation approach due to the unavailability of group IDs information. The authors of the dataset were unable to share the information due to restrictions, and therefore, we could not follow the exactly same validation approach.




\subsection{Interaction Cue Detection}
\label{converse_cue_detect}

To detect non-verbal and paralinguistic social interaction cues, we leveraged the pretrained YAMNet model \cite{YAMNet}. From these, we identified 12 classes as indicative of \emph{interaction cues}.

Prior work \cite{Liang2023-gi} focused on a narrow subset of conversation-related classes, including \emph{Conversation}, \emph{Chatter}, and \emph{Whispering}. Building on this, we retained these classes and additionally selected nine classes that capture a broader range of vocal and paralinguistic behaviors relevant to social interaction\footnote{Definitions of the 12 selected cues are available at https://research.google.com/audioset/ontology/index.html}: \emph{Speech}, \emph{Shout}, \emph{Screaming}, \emph{Laughter}, \emph{"Wail, Moan"}, \emph{Groan}, \emph{"Narration, Monologue"}, \emph{"Child Speech, Kid Speaking"}, and \emph{Clapping}.

As noted earlier, YAMNet processes audio in fixed 0.48-second frames and outputs class probability estimates for each frame across all 521 audio event classes. To derive interaction cues, we aggregated predictions from two consecutive frames (approximately one second), aligning with our FSD design, which also operates at the two-frame level (Section~\ref{fsd_section}). In contrast, assigning a single label to an entire 15-second window may obscure the presence of short but meaningful interaction cues within that interval. Accordingly, we computed the mean predicted probability across the two frames and selected the class with the highest averaged probability among all 521 classes as the final label. If this label belonged to our predefined set of 12 interaction cues, the instance was counted as containing an interaction cue.

\subsection{Social Interaction Detection Pipeline}
\label{algo_interaction_section}

\subsubsection{Recording Window and Duty Cycle.}
\label{threshold_justi_rec_window_duty_cycle_sec}

Audio sensing followed a 1.5-minute duty cycle. In each cycle, we recorded approximately 16 seconds of audio per sensing cycle, consisting of a 15-second intended recording window and an additional 1 second of padding to account for minor variability in actual recording duration (e.g., sensor initialization overhead, asynchronous callbacks, sequential sensor shutdown, and UI-thread latency). The padding increased the likelihood of obtaining at least 15 seconds of usable audio per cycle. For simplicity, we refer to this as a 15-second recording window throughout the remainder of the paper.

The choice of recording window and duty cycle was informed by prior work on conversational interaction patterns. Conversational turns are typically short, with a mean turn duration of approximately 2 seconds \cite{Levinson_2025}. Gaps between turns are often very brief, on the order of 200 ms \cite{Levinson_2016}. Based on these dynamics, we required interaction cues to be present in at least 50\% of each 15-second window. This threshold reflects the intuition that, given rapid turn-taking and short response gaps, interaction-related cues are likely to occur in a substantial fraction of windows during an ongoing interaction. Because dyadic interactions are the most common form of daily social interaction \cite{Peperkoorn_Becker_Balliet_Columbus_Molho_Van_Lange_2020}, it is reasonable to expect foreground speech to occur within at least some portion of a 15-second recording window during an ongoing interaction. At the same time, as discussed in Section~\ref{define_interact}, airtime varies substantially and individual speaking turns may extend beyond two minutes in certain settings \cite{schroeder-etal-2024-fora, Michel_Cappellini_2019}. Accordingly, we did not impose a requirement that foreground speech be present in every 15-second window. 

Our chosen duty cycle is shorter than those used in several prior systems, including the 12-minute sampling interval in the Electronically Activated Recorder (EAR) \cite{Macbeth_Bruni_De_La_Robbins_Chiarello_Montag_2022}, the 3-minute interval used in AWARE \cite{noauthor_undated-aware_framework}, and continuous audio recording employed in recent smartwatch-based systems \cite{Liang2023-gi}. One motivation for adopting a shorter duty cycle was to better capture short interactions while simultaneously improving privacy and minimizing resource consumption. For example, if a 3-minute interval similar to AWARE had been used, our system would have missed 537 interactions shorter than 3 minutes that were successfully detected during real-world deployments (Section~\ref{on_watch_perf_real_world}). On the other hand, selecting a much shorter duty cycle, such as less than 1 minute, may not be feasible in practice because the operating system can throttle frequent scheduled background tasks. Specifically, Android documentation \cite{AlarmManager} notes that, under normal system operation, such scheduled triggers are not dispatched more than about once per minute, and this interval can be substantially longer in low-power idle modes, such as around 15 minutes.

Finally, the short recording duration itself was intentionally selected to further enhance user privacy and reduce resource consumption. In comparison, the AWARE framework records 1 minute of audio per cycle \cite{noauthor_undated-aware_framework}, and EAR records 30 or 50 seconds of audio \cite{Mehl_2017_EAR}, whereas some recent systems rely on continuous recording \cite{Liang2023-gi}. By contrast, our approach limits both the duration and frequency of audio capture while retaining sufficient temporal resolution for reliable interaction detection.

Overall, the selected 1.5-minute duty cycle represents a practical balance between privacy preservation, resource consumption, and interaction detection accuracy, particularly given that prior work reports average interaction durations substantially longer than our sampling interval (e.g., approximately 5 minutes \cite{Rahman2011-kx}, and 13.97 or 20.19 minutes depending on tie strength \cite{Mastroianni_Gilbert_Cooney_Wilson_2021}).

\begin{figure}
    \centering \includegraphics[scale=0.393]{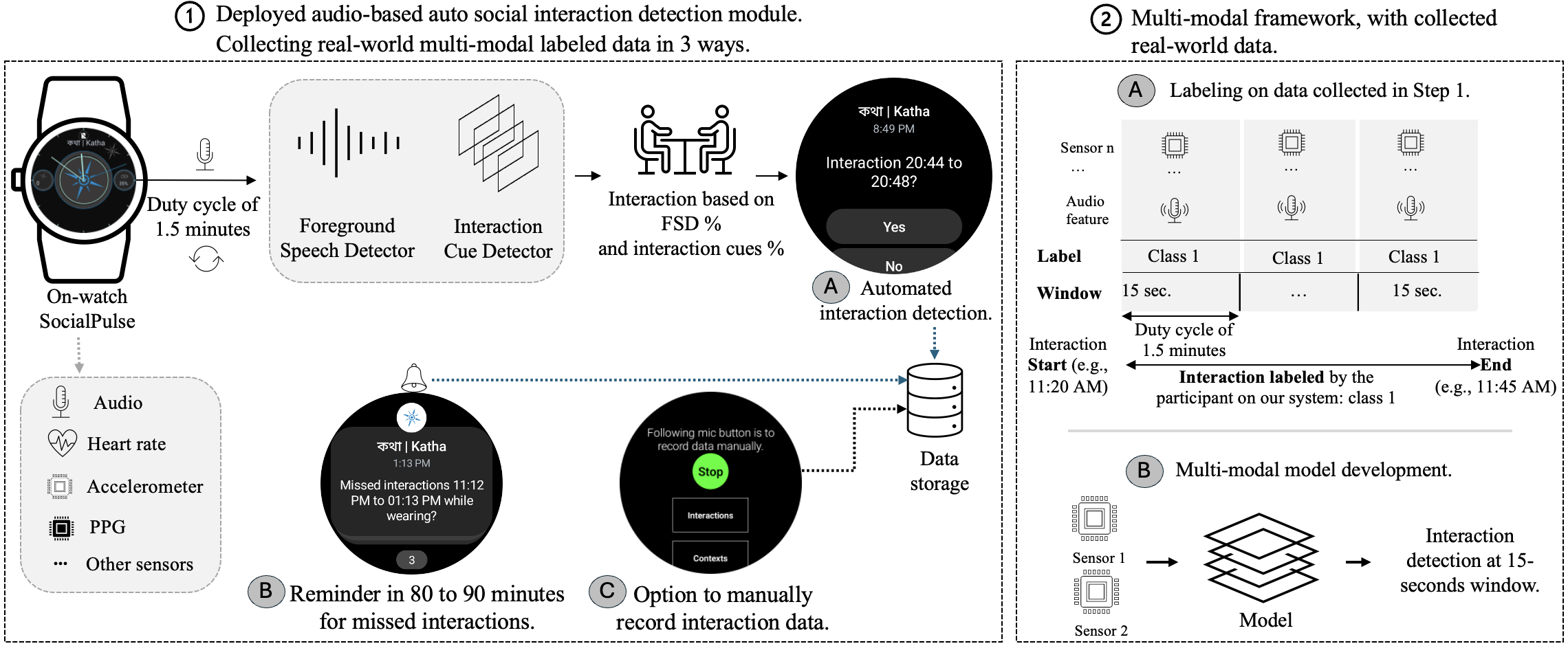}
    \caption{High-level overview of SocialPulse for automatic social interaction detection. In Step 1, real-world labeled data were collected using our deployed system, combining event-triggered prompts, periodic missed-interaction reminders, and optional manual interaction recording. In Step 2(A), labels were aligned with the 15-second data segments collected in each duty cycle. PPG data were collected continuously, while all other sensors recorded data for 15 seconds every 1.5 minutes.}
    \label{fig:high_level_overview}
\end{figure}

\subsubsection{Temporal Inference Procedure for Social Interaction Detection.}

Each 15-second audio recording is divided into non-overlapping 1-second windows for inference, just like the dataset used to train the FSD. Additionally, shorter windows reduce the likelihood of overlapping speech between the watch wearer and others within a single segment. Prior work shows that when audio is segmented into 1-second windows, mixed speech occurs only 3\% of the time in a 31-hour dataset \cite{Liang_Xu_Chen_Adaimi_Harwath_Thomaz_2023}. In contrast, longer windows (e.g., 5 seconds) are more likely to contain overlapping speech, as conversational turns last around 2 seconds on average \cite{Levinson_2025}. Similarly, interaction cues are unlikely to change at a sub-second timescale. Dividing audio into 1-second windows therefore preserves fine-grained paralinguistic cues while minimizing overlap, as emotional states typically do not fluctuate dramatically within such short intervals \cite{AlHanai_Ghassemi_2017}.  

To reduce on-device computation, we do not apply FSD to all embeddings. Instead, we first identify embeddings corresponding to interaction cues and apply the FSD only to those candidate segments, improving resource efficiency while preserving detection accuracy.


An interaction is detected when two conditions are satisfied: (1) interaction cues are present at each probe start time, and (2) the proportion of foreground speech across the full interaction exceeds our pre-defined threshold of at least 15\%. For example, if a probe starting at 3{:}20{:}00 records a 15-second segment (3{:}20{:}00--3{:}20{:}15) in which at least 50\% of the 1-second windows contain interaction cues, the system marks a potential interaction onset. If subsequent probes at 3{:}21{:}30--3{:}21{:}45 and 3{:}23{:}00--3{:}23{:}15 also satisfy this condition, the interaction is considered ongoing. If a later probe, for example, at 3{:}26{:}00--3{:}26{:}15 contains no interaction cues, the system marks the end of the interaction. If at least 15\% foreground speech is detected, the interaction is then labeled as occurring from 3{:}20{:}00 to 3{:}24{:}45, corresponding to the last probe with detected interaction cues. The full detection procedure is detailed in Algorithm~1 in the Supplementary Materials.

\section{On-Device Implementation of the Social Interaction Detection System}
\label{on_watch_imple_section}


This section focuses on the software architecture, model deployment, and runtime considerations for smartwatch-based inference. We primarily used Java for system development, including sensor control, data collection, and temporary on-device storage prior to upload to Amazon S3. Python was used for model execution, contextual data handling, and audio feature extraction since it provides a richer ecosystem and supports future system extensions across a wide range of applications. To enable interoperability between Java and Python on the smartwatch, we employed the Chaquopy Python SDK\footnote{https://chaquo.com/chaquopy/}.  

For on-device model inference, we used TensorFlow Lite, which supports low-latency execution on resource-constrained devices. To ensure compatibility with the Python environment supported by Chaquopy, we aligned library versions used during model development with those available on-device.

\subsection{Sensor Configuration and Data Collection}
\label{sensing_config}


Audio was recorded at a sampling rate of 16~kHz, which balances audio quality and power consumption relative to higher sampling rates (e.g., 44.1~kHz). Audio was initially captured in PCM format and converted to WAV format for processing using FFmpegKit \footnote{https://github.com/arthenica/ffmpeg-kit}. This design enabled efficient buffering of audio data in memory, allowing the system to access the most recent 15-second segment without re-recording. 

In addition to audio, we collected heart rate, photoplethysmography (PPG), ambient light, step count, battery status, accelerometer, and gravity sensor data. As described in Section~\ref{algo_interaction_section}, we used the same data collection window across modalities, configuring a 16-second sensing interval with an intended duration of 15 seconds. For clarity, we refer to this window as 15 seconds throughout the paper, with the understanding that it corresponds to approximately 16 seconds of data.

Among these sensors, PPG data was collected via the Samsung Health Sensor SDK\footnote{https://developer.samsung.com/health/sensor/overview.html}. PPG was sampled continuously using batched delivery, in which samples are buffered and delivered in batches rather than triggering CPU wake-ups for each event. The CPU is activated when a batch is ready, resulting in substantial power savings \cite{batching}.

All other sensors were sampled using a 1.5-minute duty cycle, consistent with the audio sensing configuration. For these sensors (except heart rate, which was collected at the fastest possible rate permitted by the hardware), we set the maximum reporting latency to 200{,}000~microseconds~\cite{monitor_sensors} to save battery consumption and on-device storage usage. Given that the system stores audio log-mel spectrograms (computed with a 25~ms window and 10~ms hop), collects PPG continuously at 25~Hz, and may operate for multiple days between participant uploads, aggressive sensor reporting could rapidly exhaust on-device memory.

At the end of each data collection window, sensing was explicitly terminated within the sensor event listeners to improve resource efficiency. Using scheduled alarms (e.g., via AlarmManager) to stop sensors at predetermined times can introduce additional battery overhead \cite{AlarmManager}. Because multiple sensors were active—and PPG data were sampled at 25~Hz—sensor callbacks were invoked frequently during each 15-second window, enabling timely termination of sensing without incurring additional scheduling overhead. To ensure high-quality data collection, the watches were configured with Always-On Display enabled to help maintain more stable sensing behavior by reducing screen-off effects particularly during resource-intensive audio processing and power management effects such as irregular sensing schedules.

\subsection{Participant Annotation of Social Interactions}
\label{all_annot_for_interact_section}

To obtain reliable ground truth for naturally occurring social interactions while minimizing participant burden, we designed a smartwatch-based annotation workflow that combines automatic detection, participant feedback, and manual reporting. Because ecological momentary assessments (EMAs) were deployed on a smartwatch with a limited screen size, all annotation questions were phrased as concisely as possible, as illustrated in Figures 2 and 3 of the Supplementary Materials. To minimize participant burden and avoid sleep disruption, notifications were suppressed during nighttime hours (12:00 AM–7:59 AM), even if participants wore the watch and potential interactions were detected.

When a notification was issued, the watch generated a 200~ms vibration to increase the likelihood that participants noticed the prompt. Participants could provide interaction annotations in three ways: (1) responding to notifications associated with automatically detected interactions, the primary mechanism, (2) manually initiating and recording an interaction, and (3) responding to periodic prompts asking whether the system missed any interactions. To account for cases in which notifications were accidentally dismissed or cleared, participants could navigate to previously issued prompts (e.g., missed interaction notifications) by selecting the corresponding option from  the app interface. 


\subsubsection{Participant Annotation of Auto-Detected Social Interactions.}

Within at most two minutes after an interaction ended, the system issued a notification (Figure~\ref{fig:high_level_overview}) asking participants whether an interaction had occurred within a given time window. Participants could respond \emph{Yes}, \emph{No}, or \emph{Maybe}. The \emph{Maybe} option was included to account for cases in which participants were uncertain or had difficulty recalling details when responding after a delay.

If the response was \emph{Yes} or \emph{Maybe}, participants reported the number of people involved, the interaction mode (e.g., in-person or virtual), and an interaction rating. An additional “?” option was provided for both the number of people involved and the interaction rating to accommodate cases in which participants could not report these values precisely. If the response was \emph{No}, participants indicated whether the detected start or end time was incorrect or whether no interaction had occurred. In such cases, participants also reported whether device-generated speech was present and whether people nearby were talking, enabling us to better understand potential sources of system errors in real-world settings.

Importantly, participants answered the same number of follow-up questions regardless of their initial response. This design choice was intended to reduce response bias, as requiring fewer questions for a particular option could incentivize its selection, particularly during busy or cognitively demanding moments.

\subsubsection{Manual Reporting of Missed Social Interactions.} 
\label{manual_interact} 

Participants periodically received a notification asking whether the system had missed any interactions for which no automatic notification had been delivered (Figure 3(b) in the Supplementary Materials). The system prompted approximately every 80 minutes, with an additional random delay of 0–5 minutes to reduce potential bias associated with fixed prompting intervals.

If participants responded \emph{Yes}, they were instructed to add all applicable missed interactions by specifying their corresponding approximate start and end times. Throughout the paper, we refer to these manually reported events as \emph{added interactions}.

\subsubsection{Manual Recording of Social Interactions.}

Participants were instructed to manually record interactions when needed, particularly during virtual interactions, where acoustic data may be unavailable (e.g., when participants wear headphones). Manual recording could be initiated and terminated by pressing a button in the watch application.

After manual recording ended, participants were prompted to answer the same set of follow-up questions used for auto-detected interactions. 

\subsection{Privacy-Preserving System Design }

Our on-device system was designed with multiple safeguards to protect participant privacy. First, all data were collected with informed consent. The consent process explicitly described the study objectives, the sensors used and the nature of the data collected. Participants were given the opportunity to ask questions and could withdraw from the study at any time without penalty. Second, sensing was automatically disabled whenever the smartwatch was removed from the participant’s wrist. To further promote transparency, our system displayed a visible ``Recording\ldots'' indicator during each 15-second window. Third, all data processing was performed on-device. Raw audio was stored only transiently and was deleted immediately after conversion into numerical representations, a process that typically completed within a few seconds. Only a single derived audio representation (the log-mel spectrogram) was retained; this representation is more privacy preserving than raw audio and is widely used in audio analysis research \cite{info_10_2196_66491}. Fourth, collected data were stored temporarily within the application sandbox, preventing access by other applications on the device, then securely uploaded to Amazon S3, which provides both client-side and server-side encryption.

Finally, all privacy and security measures were reviewed and approved by the university’s Institutional Review Board (IRB) and Information Technology Security ensuring compliance with institutional and ethical standards for human-subjects research.

\subsection{Pre-Deployment System Testing}
\label{system_testing_section}

Before the main deployment, we conducted pre-deployment testing with 11 participants to assess real-world system behavior and identify major usability or reliability issues. Participants wore the smartwatch for several days (38 participant-days total), during which the system automatically detected 343 interactions; 73.18\% ($N=251$) were confirmed as correct by participants. Follow-up discussions with six participants indicated high recall: participants did not report any interactions lasting at least two minutes that they believed occurred without a corresponding notification. Participants also reported that frequent notifications could be burdensome on socially dense days. Based on this feedback, we refined the study protocol for the main deployment by limiting participation to three days and capping interaction-related notifications at a maximum of four per hour to balance annotation quality with participant burden.

\section{In-the-Wild Deployment and Evaluation}
\label{real_world_deploy_section}

Following pre-deployment testing and system refinement, we conducted a real-world deployment to evaluate the system’s performance, usability, and robustness in everyday settings.

\subsection{Participants}

We recruited $N=38$ participants from a large public university in the southeastern United States, including undergraduate students ($n=13$), graduate students, staff, and faculty members ($n=25$). All study procedures were approved by the university’s IRB and were conducted under the supervision of a licensed clinical psychologist and members of the research team. To avoid potential bias, research assistants affiliated with the laboratories conducting the study were not eligible to participate.

The sample was predominantly composed of heterosexual Asian and White non-Latinx women, with a mean age of 25.42 years ($SD=8.19$). Participant demographics are summarized in Table~\ref{demo_info}.


\begin{table}[!htbp]
\centering
\caption{Demographic characteristics of participants in our study on 38 participants.}
\label{demo_info}
\small
\begin{tabularx}{\linewidth}{>{\raggedright\arraybackslash}X rr >{\raggedright\arraybackslash}X rr >{\raggedright\arraybackslash}X rr}
\toprule
\multicolumn{3}{c}{Gender} & \multicolumn{3}{c}{Race} & \multicolumn{3}{c}{First Language} \\
\cmidrule(lr){1-3} \cmidrule(lr){4-6} \cmidrule(lr){7-9}
Characteristic & \textit{n} & \% & Characteristic & \textit{n} & \% & Characteristic & \textit{n} & \% \\
\midrule
Woman & 20 & 52.63 & Caucasian/White & 14 & 36.84 & English & 25 & 65.79 \\
Man & 17 & 44.74 & Asian & 17 & 44.74 & Not English (e.g., Bengali) & 13 & 34.21 \\
Non-Binary & 1 & 2.63 & Multiple Races & 4 & 10.53 &  &  &  \\
 &  &  & African American & 2 & 5.26 &  &  &  \\
 &  &  & Middle Eastern & 1 & 2.63 &  &  &  \\
\bottomrule
\end{tabularx}
\end{table}

\subsection{Study Design and Participant Protocol}

Participants were recruited through two channels: (1) a university research participation pool for paid research studies and (2) targeted advertisements distributed via email lists within specific schools, departments and research labs. 

After enrollment, participants completed an initial study visit that included informed consent and hands-on instruction on how to use our system (e.g., uploading data and responding to ecological momentary assessments). Participants were provided with a detailed user manual during the visit and received an electronic copy by email afterward. Study visits were conducted by trained research staff and undergraduate research assistants affiliated with a psychology laboratory. 

Each participant was instructed to wear the smartwatch for three consecutive days, including two weekdays and one weekend day (either Thursday–Saturday or Sunday–Tuesday). This schedule was designed to capture variability in social interactions across structured weekdays and less-structured weekends. 

Researchers monitored participant compliance with smartwatch wearing and EMA completion throughout the study, sending reminder emails as needed. Participants could earn up to \$50 in compensation, with payment determined by participation rate, calculated as the product of smartwatch wear duration and the proportion of EMA prompts completed within two hours.

This project was automatically issued a National Institutes of Health (NIH) Certificate of Confidentiality upon grant award. All study data were de-identified prior to analysis and sharing in accordance with funder guidelines and open science best practices.

\subsection{Data Quality and Sensor Coverage}
\label{data_quality_section}

Across the deployment, participants wore the smartwatch for a total of 120 days, with an average usage duration of 3.16 days per participant ($SD = 0.64$). Though participants were instructed to wear the smartwatch for three consecutive days, some participants wore the device for longer periods. In such cases, we included up to four days of data. For simplicity, we refer to the study duration as three days throughout the paper, reflecting the average usage of 3.16 days.

The smartwatch collected nine total data streams: Audio, PPG, Accelerometer, Gravity, Light, Heart rate, Off-body, Step Count, and Battery Level. For model development, we excluded Heart Rate, Step count, and Battery sensors. Instead of Heart Rate, we relied on PPG data, which captures beat-to-beat interval dynamics and thus offers richer physiological information than heart rate expressed in beats per minute. We excluded Step Count and Battery Level because they were recorded once per probe start, and limited variability in these sensors was expected within a 15-second sensing window.

\begin{figure}[!htbp]
    \centering
    \includegraphics[width=1\linewidth]{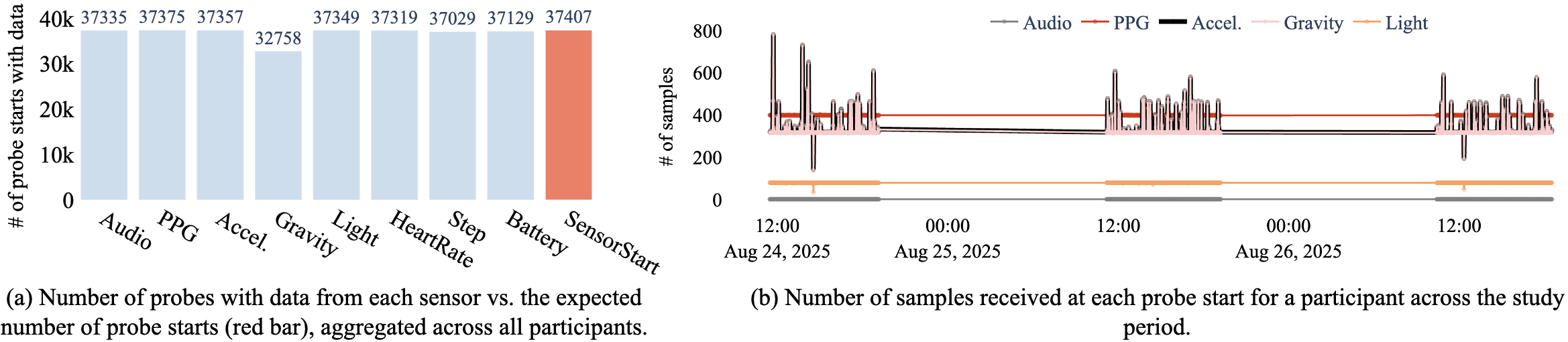}
    \caption{Sensing data quality of the on-device system in real-world deployment. 
    (a) Sensor activation counts across participants and days. 
    (b) Sample counts for sensors used in multimodal analysis. Heart rate, step, and battery sensors are excluded for clarity. The accelerometer line width is increased to distinguish it from the gravity sensor, which has a similar sampling rate.
}
    \label{fig:sensing_data_quality}
\end{figure}

As shown in Figure~\ref{fig:sensing_data_quality}(b), for the sensors used in multimodal model development (Section~\ref{multi_modal_section}), the number of collected samples was largely consistent across the study period, with greater variability observed for the Accelerometer and Gravity sensors. Despite this variability, when data were present, observed sample counts were often higher than the nominal sampling rates.

One plausible explanation for these elevated sampling rates is the Android-based Wear OS sensor architecture, in which multiple applications communicate with the Android sensor framework and share a single-client sensor hardware abstraction layer (HAL) \cite{sensor_stack}. When different applications request the same sensor at different sampling rates (e.g., one at 10~Hz and another at 20~Hz), the sensor delivers data at the highest requested rate. As a result, applications requesting lower rates may still receive data at higher frequencies \cite{sensor_stack}. Consequently, sensing data cannot be guaranteed to arrive at a perfectly constant sampling rate.

During the deployment period, all sensors were activated more than 37{,}000 times for data collection, closely matching the expected number of sensor activations (Figure~\ref{fig:sensing_data_quality}(a)), with the exception of the Gravity sensor.


\subsection{In-the-Wild System Performance}
\label{on_watch_perf_real_world}

\subsubsection{EMA Response Latency and Missed Interactions.}
\label{ema_response_latency}

Participants responded promptly to event-triggered EMA notifications, with 60.14\% of responses occurring within 1 minute and 90.96\% within 2 hours, supporting the reliability of the collected annotations (Section 1 of Supplementary Materials). Based on pre-deployment testing, interaction-related notifications were capped at four per hour to balance annotation coverage with participant burden, resulting in 232 automatically detected interactions not being sent for annotation.

Missed-interaction prompts indicated that participants reported no missed interactions in the majority of cases, with relatively few added or edited interactions. Analysis of these cases suggests that discrepancies may arise from short-duration interactions, virtual interactions, or periods when the watch was not continuously worn. A detailed breakdown of missed and participant-edited interactions is provided in the Section 2 of Supplementary Materials.

\begin{figure}[!htbp]
    \centering
    \includegraphics[width=1\linewidth]{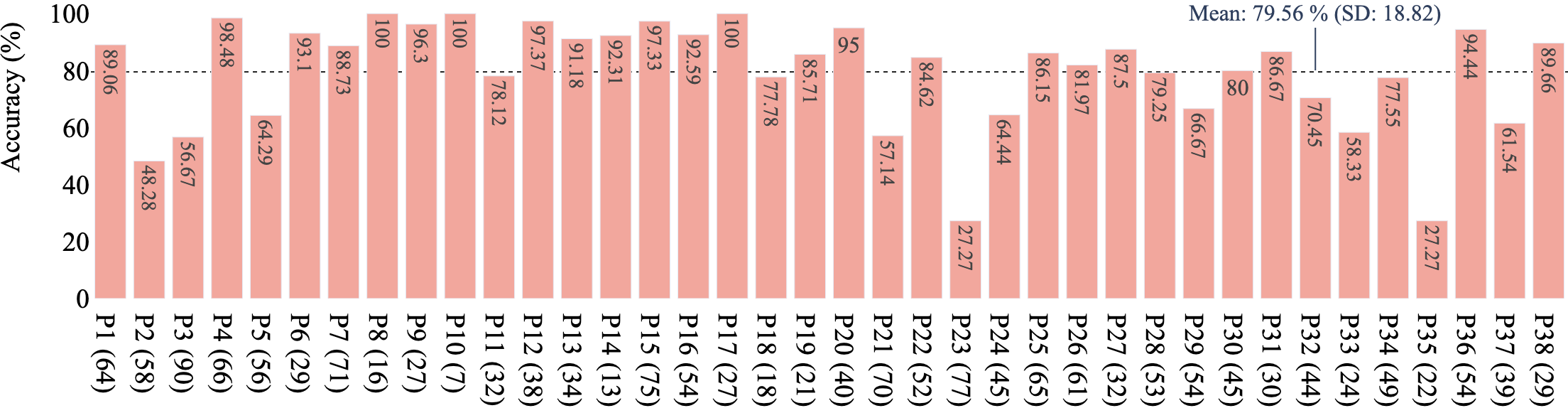}
    \caption{Per-participant accuracy of the deployed auto-detection system. Values in parentheses indicate the total number of automatically detected interactions to which each participant responded. The dotted black line indicates the mean accuracy across all participants.}
    \label{fig:accuracy_per_p}
\end{figure}

\subsubsection{Performance of the Deployed System in the Wild.}
\label{eval_of_deployed_system_section}

Among the 1{,}691 automatically detected interactions, ten were labeled as \emph{maybe}, indicating participant uncertainty about whether an interaction occurred, consistent with the definition provided in the user manual. These ten cases were excluded from performance evaluation. Of the remaining interactions, 77.28\% ($N=1{,}299$) were confirmed as correct by participants. The F1 score was 84.87\%. To ensure consistency in the precision and recall calculations for F1, we excluded all overlapped interactions.

Participant-level analysis showed a mean accuracy of 79.56\% ($SD=18.82$\%), with per-participant accuracy ranging from 27.27\% to 100\% (Figure~\ref{fig:accuracy_per_p}). Notably, for 71.05\% of participants, the system achieved an accuracy of at least 75\%.

\begin{figure}[!htbp]
    \centering
    \includegraphics[width=1\linewidth]{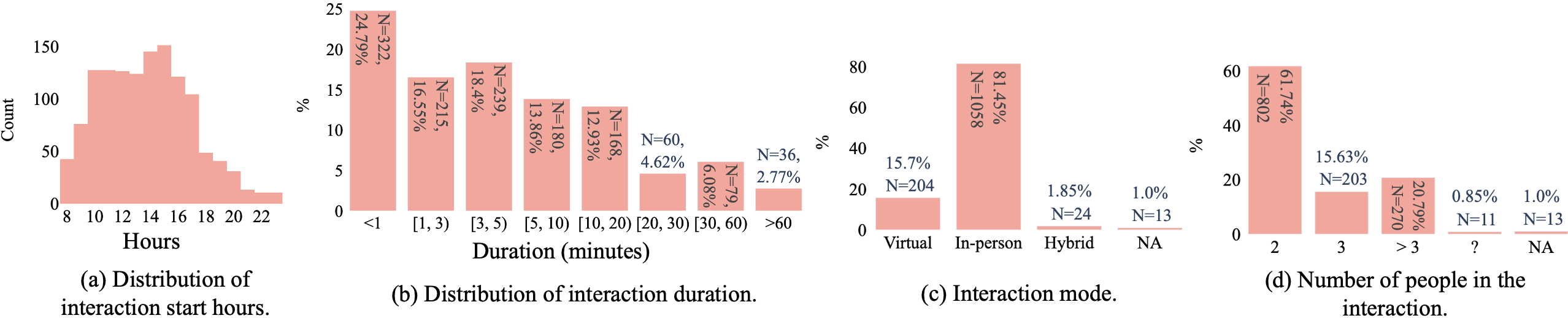}
    \caption{Statistics on correctly auto-detected interactions in real-time by our on-watch system. NA: Not available. "?" denotes "Do not know."}
    \label{fig:stat_correct_interact}
\end{figure}

Analysis of correctly detected interactions indicated that interaction onsets occurred across hourly bins from 8:00~AM to 11:00~PM (Figure~\ref{fig:stat_correct_interact}(a)). Interaction durations ranged from under 1 minute ($N=322$, 24.79\%) to over 1 hour ($N=36$, 2.77\%) (Figure~\ref{fig:stat_correct_interact}(b)). All interaction modes were observed, including in-person ($N=1{,}058$, 81.45\%), virtual ($N=204$, 15.70\%), and hybrid ($N=24$, 1.85\%) interactions (Figure~\ref{fig:stat_correct_interact}(c)). Most interactions involved two participants ($N=802$, 61.74\%), as reported by participants (Figure~\ref{fig:stat_correct_interact}(d)).

\subsubsection{Analysis of Social Interaction Detection Errors.}

\begin{figure}
    \centering
    \includegraphics[width=1\linewidth]{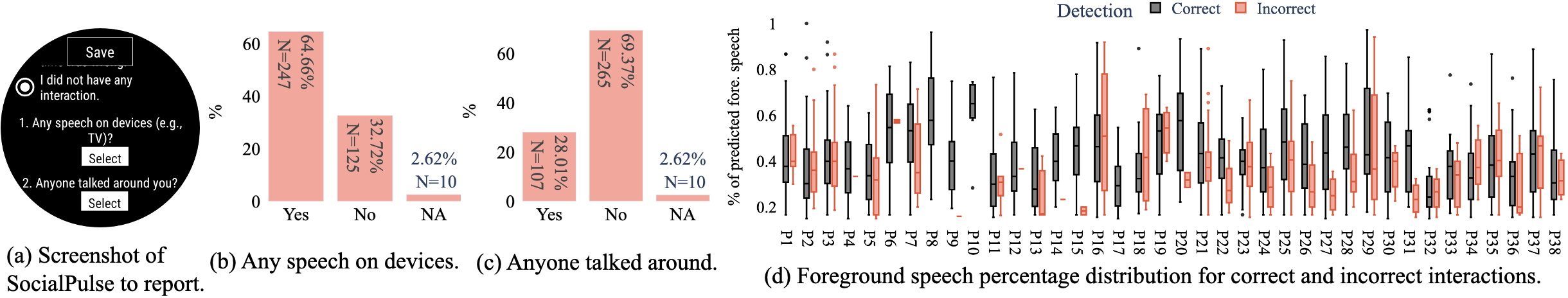}
    \caption{(a) Screenshot of our reporting interface, when a participant marked an interaction as inaccurate. (b–c) Distributions of participant responses related to surrounding speech. (d) Distribution of predicted foreground speech percentages for correct and incorrect interactions. NA: Not Available.}
    \label{fig:incorrect_detection_auto}
\end{figure}

To investigate sources of detection errors, participants reported whether speech from nearby devices or nearby other people was present during incorrectly detected interactions. Among incorrect detections, 64.66\% ($N=247$) involved speech from electronic devices, while 28.01\% ($N=107$) involved speech from nearby people (Figure \ref{fig:incorrect_detection_auto}), indicating that non-interaction speech remains a primary source of confusion in real-world settings.

To quantify differences between correct and incorrect detections, we analyzed the proportion of foreground speech within detected interactions. For most participants, foreground speech percentages were higher in correctly detected interactions than in incorrect ones. We compared per-participant mean foreground speech percentages using a Wilcoxon signed-rank test excluding three participants with 100\% accuracy. Across the remaining 35 participants, the Wilcoxon test revealed a statistically significant difference ($p=0.00009$), suggesting that participant-specific foreground speech thresholds, rather than the fixed 15\% threshold used in the deployed system, may further reduce errors.

Despite this robustness, some errors persisted, particularly when speech-emitting devices or nearby conversations were in close proximity to the watch. The relatively low foreground speech threshold of 15\% was intentionally selected to prioritize sensitivity and reduce missed interactions, supporting higher-quality annotation at the expense of some false positives. These findings highlight opportunities for adaptive or personalized thresholds in future systems.

\begin{figure}[!htbp]
    \centering
    \includegraphics[width=1\linewidth]{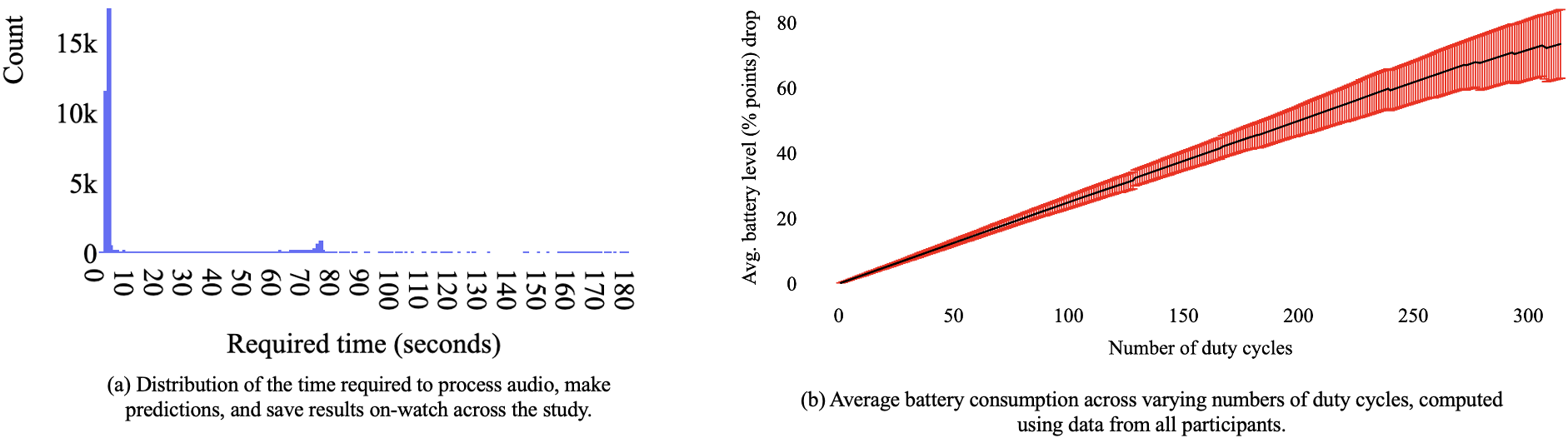}
    \caption{Resource consumption of SocialPulse system. (a) Processing time per probe, computed from 37{,}353 processing-time logs collected across the study. For readability, only instances with processing times of at most 3 minutes are shown, accounting for 99.84\% of all recorded probes. In b, the black line represents the average battery consumption per participant over m duty cycles, with each duty cycle set to 1.5 minutes. The red error bars represent the 95\% confidence intervals.}
    \label{fig:resource_consump}
\end{figure}

\subsubsection{On-Device Resource Consumption.}
\label{resource_consump_sec}

Across 37{,}353 probe start events, we recorded the time required to process audio, perform inference, and save outputs (Figure~\ref{fig:resource_consump}(a)). The mean processing time per probe was 13.286 seconds ($SD = 27.28$), while the median was substantially lower at 3.183 seconds. In 77.87\% of probes, total processing time was under 4 seconds, indicating that most inference operations completed quickly despite occasional longer delays.

We also analyzed power consumption during deployment, based on data of all participants. To account for variation in initial battery levels across participants and days, we segmented the battery data. A new segment was defined when the time gap between consecutive samples at least 2 minutes (suggesting the watch was removed) or when the battery level increased (indicating charging). Within each segment, we computed battery-level drop over $m$ duty cycles. Figure~\ref{fig:resource_consump}(b) reports battery consumption for up to 314 duty cycles, corresponding to the average number per participant per day (313.7). On average, 314 duty cycles consumed 73.51\% of the battery, which corresponds to approximately 7.85 hours under the configured 1.5-minute duty cycle.

Multiple system components likely contributed to overall energy consumption. These include: (1) the duty-cycle scheduler, which triggers sensing and processing every 1.5 minutes; (2) audio recording, on-device processing, model loading, and inference, which together constitute a substantial computational load; (3) cross-runtime execution via the Chaquopy SDK, which introduces overhead due to Java–Python transitions; and (4) continuous physiological sensing, particularly PPG. During deployment, participants used dedicated study watches and were expected not to use them beyond study-related tasks (e.g., responding to EMAs and uploading data).




\section{Exploration of Multimodal Model for Social Interaction Detection}
\label{multi_modal_section}

Building on the real-world deployment and participant-provided annotations, we next explored a multimodal model to identify social interactions using synchronized signals across modalities, enabling the capture of interaction dynamics beyond any single sensor stream. One limitation of the deployed framework is that it determines whether an interaction occurred within a candidate interval only after the interval has ended, based on the proportion of foreground speech and the presence of interaction cues within each 15-second window. This design constrains applications that require more timely detection, such as adaptive sensing (e.g., initiating richer data collection at interaction onset in studies of social anxiety) and context-aware systems.

To address this limitation, we trained models at the 15-second window level using labels derived from the real-world deployment data. This formulation was not feasible prior to deployment, as labeled real-world interaction data were not yet available. At that stage, we only had access to foreground speech labels collected in a semi-controlled setting, which supported foreground speech detection but not interaction-level modeling.

\subsection{Labeling Strategy for Multimodal Model Training}

To train the multimodal interaction detection model, we assigned labels to each 15-second data collection window based on participant-confirmed interaction annotations (Figure~\ref{fig:high_level_overview}). Windows whose timestamps fell within participant-labeled interaction intervals were labeled as class~1 (presence of social interaction). Windows outside these intervals were treated as candidates for class~0 (no interaction), subject to conservative filtering described below.

Results from the deployed interaction detection system and participant responses to missed-interaction notifications indicate high recall (see details in Section 2 of Supplementary Materials). That is, when an interaction occurred, the system was highly likely to notify the participant. This observation motivated us to include windows with no detected interactions as class~0, provided there was sufficient evidence that no interaction occurred during those intervals.

Reliable ground-truth labels were unavailable for automatically detected interaction intervals when participants did not respond to notifications. To preserve labeling accuracy, we excluded all windows associated with such unresponded detections from model training.

Because raw audio was not stored for privacy, we reconstructed the deployed interaction detection pipeline offline using the on-device–saved log-mel spectrogram features. This reconstruction enabled us to identify excluded interaction intervals and verify labeling consistency without requiring access to raw audio. We confirmed that the reconstructed interaction boundaries aligned with those produced by the deployed system within the 15-second sensing resolution used for labeling.

To further reduce the risk of mislabeling non-interaction windows, we applied conservative filtering informed by exploratory analysis of foreground speech. Specifically, we excluded candidate interaction windows with ambiguous foreground speech levels, retaining only windows with clear evidence of interaction or non-interaction. Given that such ambiguous cases were rare and labeling was performed at a 15-second resolution, we do not expect these exclusions to affect model performance or generalization. In addition, we provide detailed analysis in Section 3 of the Supplementary Materials on the fine-grained temporal reliability and label quality of participant-added interactions reported in response to missed-interaction notifications.

\subsection{Multimodal Model Design and Training}

We developed a convolutional neural network (CNN)–based multimodal model to identify social interactions from wearable sensing data. CNNs are well suited for capturing complex patterns in structured representations and are more parameter-efficient than fully connected networks due to weight sharing and partial connectivity \cite{ml_bookAurélien}. Because each probe captured only 15 seconds of data, the number of raw samples per modality was limited (e.g., approximately 90 accelerometer samples per probe). Such short time-domain signals are often insufficient for training deep learning models directly. We therefore transformed all raw sensor signals into spectrogram-based representations.

\paragraph{Sampling Rate Normalization.}
 To ensure consistent representations across all probe windows, we resampled each sensor stream to a fixed sampling rate per modality. For each sensor, we selected the most frequently observed sampling rate and rounded minor variations. For example, accelerometer sampling rates of 6.125~Hz, 6.25~Hz, and 6.3125~Hz were treated as 6~Hz and resampled accordingly. When a probe window contained more samples than required, we applied averaging; when samples were missing, we used interpolation. Each probe window was processed independently to avoid information leakage and to reflect real-world inference conditions.

An exception was made for the gravity sensor. Although its most frequent sampling rate was 19.8125~Hz, the second most frequent rate was 6.25~Hz. Using the higher rate would have required extensive interpolation for more than 5{,}000 probe windows sampled at approximately 6~Hz, potentially distorting the signal. We therefore adopted a 6~Hz sampling rate for the gravity sensor to minimize missing data and better preserve real-world signal characteristics.

\paragraph{Feature Construction.}
For accelerometer and gravity signals, we computed the magnitude across the three axes. PPG signals were preprocessed using NeuroKit2 \cite{Makowski2021neurokit}, and spectrograms were generated from the cleaned waveforms. To balance time and frequency resolution given the limited data per probe, we used a segment length of 16 samples, an FFT size of 16, and an overlap of 12 samples. Based on the resulting sample counts, spectrogram images of size $112 \times 112$ were used. During preliminary experiments, smaller images consistently outperformed larger ones (e.g., $112 \times 112$ vs.\ $224 \times 224$), likely because compact representations better preserve informative structure given the limited signal length. All reported results use $112 \times 112$ images.

\begin{figure}[htbp]
    \centering
    \includegraphics[width=0.98\linewidth]{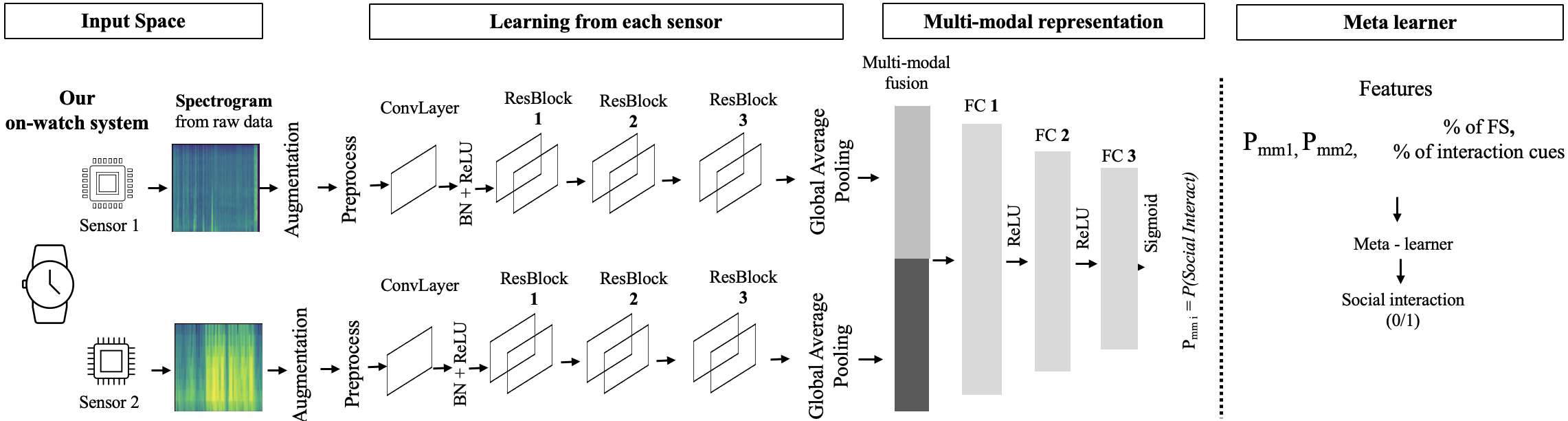}
    \caption{Architecture of our multi-modal model for 15-second window-level prediction. ConvLayer: Convolution layer, ResBlock: Residual block, FC: fully connected. \(P_{mm1}\) and \(P_{mm2}\) presents the probability to have an interaction based on multi-modal model 1 (e.g., Audio-Accelerometer) and multi-modal model 2 (e.g., Audio-PPG) respectively. FS: Foreground speech.}
    \label{fig:multi_modal_arch}
\end{figure}

\paragraph{Model Architecture.}
The multimodal architecture is shown in Figure~\ref{fig:multi_modal_arch}. Each modality was processed by a convolutional layer followed by batch normalization and ReLU activation, then three residual blocks \cite{He_2016_CVPR}. Skip connections in the blocks facilitate optimization and improve representational learning \cite{ml_bookAurélien}. We used He initialization, consistent with ResNet-style architectures. We did not adopt a full ResNet model due to its large parameter count (e.g., $\sim$11.7 million parameters for ResNet-18), which is impractical for real-time on-watch inference.

After the residual blocks, we applied global average pooling (GAP) to each modality-specific branch. The resulting feature vectors were concatenated to form a fused multimodal representation, followed by two fully connected layers with 64 and 32 neurons (ReLU activation). A final sigmoid-activated output layer produced the probability of social interaction.

To reduce overfitting, we applied on-the-fly data augmentation during training, including temporal flipping, slight temporal stretching (scale=1.001–1.05) followed by cropping to $112 \times 112$, and additive Gaussian noise (mean=0, SD=0.05). Stretching was applied along the temporal axis to preserve meaningful temporal dynamics in the spectrograms.

\paragraph{Training Procedure.}
Because social interactions are less frequent than non-interaction periods, we used weighted binary focal loss to address class imbalance. The model was optimized using stochastic gradient descent with a learning rate of 0.05, batch size of 32, and up to 50 epochs. Early stopping was applied based on validation loss, terminating training if the loss did not decrease for five consecutive epochs. Training data were shuffled prior to learning to avoid ordering artifacts.

\paragraph{Cross-Validation Strategy.}
We evaluated the model using a Leave-One-Participant-Out Cross-Validation (LOPOCV) scheme. In each iteration, one participant was held out for testing. From the remaining $n-1$ participants, two with the smallest class-count difference were selected for validation to reduce bias from extreme class imbalance. The remaining $n-3$ participants were used for training.

\paragraph{Meta-Learning Integration.}
In a final meta-learning stage, we incorporated the percentage of interaction cues and the percentage of foreground speech as additional features, together with predicted probabilities of social interaction from two multimodal models (e.g., Accelerometer–Audio and Light–Audio). Foreground speech was estimated using the FSD (Section~\ref{fsd_section}) only for frames containing interaction cues, consistent with the deployed system (section \ref{algo_interaction_section}). Following the criterion used in the deployed framework, FSD was triggered only when interaction cues were present in at least 50\% of a 15-second window. To evaluate this design choice, we conducted a sensitivity analysis by varying the interaction-cue threshold from 0\% to 100\% in 10\% increments (Figure~4 in the Supplementary Materials), which confirmed that 50\% yields the optimal performance. We evaluated the top-performing meta-learners from the FSD analysis—Logistic Regression, LinearSVC, and NearestCentroid — favoring these simple models for their generalization ability and reduced risk of overfitting.

\subsection{Model Evaluation and Baseline Models}

To evaluate model performance, we report accuracy, precision, sensitivity, specificity, and F1 score. Precision and F1 score are computed using macro averaging, which assigns equal weight to each class. We additionally report balanced accuracy, defined as the average of sensitivity and specificity. Balanced accuracy and sensitivity are treated as the primary evaluation metrics, as they better capture performance under class imbalance, with sensitivity reflecting the model’s ability to correctly identify interaction windows.

As baseline models for interaction detection, we used MobileNetV1 \cite{mobile_nets} and YAMNet \cite{YAMNet}, two publicly available architectures that are widely used across mobile and audio-based prediction tasks. MobileNetV1 contains approximately 3.3 million parameters and is optimized for mobile and embedded devices. We adapted the baseline development procedures described in \cite{Liang2023-gi}, which also evaluated these architectures for interaction-related tasks.

For YAMNet, we implemented the ``YAMNet + FC'' baseline from \cite{Liang2023-gi}, in which three fully connected layers are trained on YAMNet embeddings. Specifically, embeddings were averaged across each 15-second window, and the resulting vector was used as input to the fully connected classifier. For MobileNetV1, we trained the model from scratch, adapting the general approach described in \cite{Liang2023-gi}.

When implementation details were not explicitly specified in \cite{Liang2023-gi}, we followed standard practice. For example, the prior work does not describe how validation and test sets were separated during training. Therefore, consistent with prior studies using LOPOCV \cite{Kwon_Tong_Haresamudram_Gao_Abowd_Lane_Plötz_2020}, we selected one participant for validation in each fold when training baseline models. The validation participant was randomly chosen in each iteration, as using the same participant for validation across all LOPOCV folds could introduce bias.

We could not implement the main architecture proposed by \cite{Liang2023-gi} as baselines for two practical reasons. First, we needed raw audio to generate spectrograms following their feature-extraction procedure. Second, their fusion design relies on constructing spectrograms from 30 consecutive 1-second sub-clips within each 30-second window; without raw audio, we could not generate spectrograms at this 1-second granularity.

\subsection{Performance of the Multimodal Model}
\label{perf_multi_mod_section}

Across all evaluated configurations, multimodal models that included the audio modality consistently outperformed models based solely on non-acoustic sensors (Table~\ref{multi_modal_1_modal_perf}). The best-performing configuration combined accelerometer and audio data, achieving a balanced accuracy of 86.18\%. Other audio-inclusive combinations (e.g., Gravity–Audio, Light–Audio, and PPG–Audio) also demonstrated strong performance, with balanced accuracies of 85.75\%, 85.67\%, and 85.65\%, respectively.

In contrast, models relying exclusively on non-acoustic modalities performed substantially worse. For example, the PPG–Accelerometer configuration achieved a balanced accuracy of approximately 51\%, close to chance level. These results highlight the importance of acoustic information for reliable social interaction detection, even when combined with rich physiological and inertial sensing data.

\begin{table}[htbp]
\centering
\begin{tabular}{lccccccc}
\toprule
    Sensors & N samples & Precision & Sensitivity & Specificity & BA & F1 & Accuracy \\
    \midrule
    \textbf{Accelerometer\_Audio} & \textbf{33700} & \textbf{83.97} & \textbf{85.36} & \textbf{87.00} & \textbf{86.18} & \textbf{84.86} & \textbf{86.49} \\
    Accelerometer\_Gravity & 29624 & 55.19 & 13.19 & 91.26 & 52.22 & 49.72 & 67.50 \\
    Accelerometer\_Light & 33709 & 59.48 & 9.87 & 95.33 & 52.60 & 48.49 & 68.43 \\
    Gravity\_Audio & 29621 & 83.25 & 84.80 & 86.71 & 85.75 & 84.25 & 86.13 \\
    Light\_Audio & 33704 & 83.47 & 84.68 & 86.65 & 85.67 & 84.35 & 86.03 \\
    Light\_Gravity & 29628 & 55.53 & 7.68 & 95.14 & 51.41 & 46.86 & 68.53 \\
    PPG\_Accelerometer & 33700 & 52.55 & 12.06 & 90.16 & 51.11 & 48.14 & 65.57 \\
    PPG\_Audio & 33683 & 83.30 & 85.06 & 86.25 & 85.65 & 84.23 & 85.87 \\
    PPG\_Gravity & 29613 & 54.31 & 14.86 & 89.41 & 52.13 & 50.13 & 66.72 \\
    PPG\_Light & 33694 & 49.95 & 6.18 & 93.80 & 49.99 & 44.75 & 66.21 \\

    \midrule
    Accelerometer & 33727 & 46.91 & 3.31 & 95.57 & 49.44 & 42.75 & 66.53 \\
    \textbf{Audio} & \textbf{33858} & \textbf{84.53} & \textbf{84.85} & \textbf{87.98} & \textbf{86.41} & \textbf{85.33} & \textbf{87.00} \\
    Gravity & 29639 & 51.06 & 8.37 & 92.35 & 50.36 & 46.39 & 66.80 \\
    Light & 33721 & 53.27 & 5.11 & 96.13 & 50.62 & 44.60 & 67.48 \\
    PPG & 33752 & 47.77 & 8.07 & 90.20 & 49.13 & 45.04 & 64.34 \\
    \bottomrule
    
\end{tabular}
\caption{Performance of multimodal and single modal models without the meta-learner. The highlighted Accelerometer\_Audio row indicates the best-performing multimodal model in terms of balanced accuracy (BA), while the highlighted Audio row indicates the best-performing single-sensor model. N samples presents the number of samples used for model evaluation.}
\label{multi_modal_1_modal_perf}
\end{table}

After evaluating the multimodal models, we developed a meta-learner as defined in the multimodal architecture (Figure~\ref{fig:multi_modal_arch}). The meta-learner integrates the predicted probabilities from two multimodal models along with two summary features computed for each data collection window: the percentage of foreground speech and the percentage of detected interaction cues.

Across different pairs of multimodal models, the meta-learner achieved comparable performance, with balanced accuracy consistently exceeding 90\%, indicating robustness to the choice of input models. We report results using the Accelerometer--Audio and Light--Audio configurations, which correspond to the top-performing multimodal models (Table~\ref{multi_modal_1_modal_perf}), excluding Gravity--Audio. Although the Gravity--Audio configuration achieved a slightly higher balanced accuracy than Light--Audio, the Light--Audio modality provided a substantially larger number of samples for evaluation, enabling more reliable performance estimates.

Using these modalities, the LinearSVC-based meta-learner achieved a sensitivity of 91.17\%, a specificity of 89.56\%, and a balanced accuracy of 90.36\%, outperforming all baseline models.


\begin{table}[htbp]
\centering
    \begin{tabular}{lccccccc}
        \toprule
        Algorithm & N & Precision & Recall & Specificity & BA & F1 & Accuracy \\
        \midrule
        Multi-modal meta (Logit) & 33698 & 88.02 & 90.59 & 90.00 & 90.30 & 88.97 & 90.19 \\
        Multi-modal meta (NearestCentroid) & 33698 & 88.18 & 90.38 & 90.27 & 90.33 & 89.09 & 90.31 \\
        \textbf{Multi-modal meta (LinearSVC)} & \textbf{33698} & \textbf{87.86} & \textbf{91.17} & \textbf{89.56} & \textbf{90.36} & \textbf{88.88} & \textbf{90.07} \\
        \midrule
        Audio only meta (Logit) & 33698 & 88.17 & 90.54 & 90.20 & 90.37 & 89.10 & 90.31\\
        Audio only meta (NearestCentroid) & 33698 & 88.15 & 90.61 & 90.15 & 90.38 & 89.09 & 90.29 \\
        \textbf{Audio only meta (LinearSVC)} & \textbf{33698} & \textbf{87.97} & \textbf{91.01} & \textbf{89.77} & \textbf{90.39} & \textbf{88.97} & \textbf{90.16} \\
        \bottomrule
        Baseline 1 (MobileNetV1) & 33698 & 82.93 & 71.18 & 91.07 & 81.13 & 81.92 & 84.81 \\
        Baseline 2 (YAMNet + FC) & 33698 & 87.77 & 82.36 & 92.57 & 87.47 & 87.62 & 89.36 \\
        \bottomrule
    \end{tabular}
\caption{Performance of different algorithms used to train our meta-learner. The multi-modal meta-learners were trained with predicted probabilities of Accelerometer\_Audio and Light\_Audio based multi-modal models, along with percentage of foreground speech and interaction cues. The audio-only meta-learner was trained using predicted probability from the Audio-only model, together with foreground speech and interaction cues. Highlighted rows indicate the best-performing multimodal and audio-only meta-learners in terms of balanced accuracy (BA). N presents the number of samples.}
\label{meta_learners_perf}
\end{table}

\subsection{Single-Sensor Ablation and Performance of the Audio-Only Model}
\label{audio_perf_on_15_sec_windows}

To identify the contribution of individual sensors and model components, we conducted an ablation study. Specifically, we trained single-sensor models using the same architecture as the multimodal model, restricting the network to a single sensor branch (i.e., retaining only the residual blocks corresponding to one sensor in Figure~\ref{fig:multi_modal_arch} and removing all other modality-specific components).

Consistent with the multimodal results, the audio-only model substantially outperformed models based on other individual sensors. The audio-only configuration achieved a balanced accuracy of 86.41\%, whereas the accelerometer-only model achieved a balanced accuracy of 49.44\% (Table~\ref{multi_modal_1_modal_perf}). This finding underscores the dominant role of acoustic information in distinguishing interaction from non-interaction in naturalistic settings. Notably, the audio-only model contains only 0.315 million parameters, compared with 0.63 million parameters for the multimodal model, indicating that strong performance can be achieved with a relatively lightweight configuration.

We further evaluated whether the audio-only model could benefit from a meta-learner. The meta-learner incorporated the predicted probability from the audio-only model, along with the percentage of foreground speech and interaction cues, both derived from audio. For a consistent comparison with the multimodal meta-learner, we first evaluated the audio-only meta-learner on the same 33{,}698 windows. In this setting, LinearSVC achieved the best performance, with a sensitivity of 91.01\% and a balanced accuracy of 90.39\%, closely matching the performance of the multimodal meta-learner (Table~\ref{meta_learners_perf}).

We also evaluated the audio-only meta-learner on all 33{,}858 available audio windows and observed similar performance. In this case, NearestCentroid performed best, achieving a balanced accuracy of 90.37\%, sensitivity of 90.60\%, specificity of 90.13\%, precision of 88.12\%, and an F1 score of 89.06\%.

These findings suggest that performance comparable to the multimodal model can be achieved using the audio modality alone.



\subsection{Ablation Study of the Deep Learning Architecture}
To further examine the contribution of individual architectural components, we conducted an ablation study by systematically removing components from the network, ranging from the ResNet-based preprocessor to the fully connected layers (Table \ref{ablation_dl}). When excluding the ResNet-based preprocessor, we applied standard feature scaling, as neural networks typically require normalized inputs. Under this configuration, the Accelerometer--Audio model exhibited an approximately 1\% reduction in both sensitivity and balanced accuracy. Given that the evaluation was performed on 33{,}700 samples, this performance decrease is non-trivial.

Similar degradations were observed when removing most other components. Although excluding the second fully connected layer resulted in a marginally higher balanced accuracy (86.31\% vs.\ 86.18\%) for the Accelerometer--Audio model, this change was accompanied by slight reductions in F1 score and precision. In contrast, for the audio-only model, removing the second fully connected layer led to an approximately 1\% decrease in balanced accuracy, highlighting the utility of this component for audio-based interaction detection.

Overall, these results suggest that while the proposed architecture is effective, certain components may be removed to reduce model complexity when parameter efficiency is prioritized, with only modest trade-offs in performance.


\begin{table}[htbp]
\centering
    \begin{tabular}{lcccccccccccc}
        \toprule

        & \multicolumn{6}{c}{\textbf{Accelerometer and Audio}} & \multicolumn{6}{c}{\textbf{Audio}} \\
        \cmidrule(lr){2-7}\cmidrule(lr){8-13}
        
        Config.
        & N & Prec. & Sens. & Spec. & BA & F1
        & N & Prec. & Sens. & Spec. & BA & F1 \\
        
        \midrule

        e\_aug & 33700 & 83.02 & 84.36 & 86.19 & 85.28 & 83.92 & 33858 & 84.92 & 83.88 & 88.85 & 86.37 & 85.56 \\
        e\_ResNetPro & 33700 & 82.75 & 83.65 & 86.15 & 84.90 & 83.61 & 33858 & 82.85 & 82.68 & 86.74 & 84.71 & 83.63 \\
        \midrule
        
        e\_Conv & 33700 & 83.30 & 84.45 & 86.53 & 85.49 & 84.18 & 33858 & 84.08 & 85.93 & 86.93 & 86.43 & 85.02 \\
        \midrule
        
        e\_1st\_resB & 33700 & 83.83 & 84.59 & 87.17 & 85.88 & 84.68 & 33858 & 83.01 & 83.12 & 86.76 & 84.94 & 83.81 \\
        e\_mid\_resB & 33700 & 83.43 & 85.80 & 86.09 & 85.94 & 84.40 & 33858 & 83.18 & 85.11 & 86.08 & 85.60 & 84.12 \\
        e\_last\_resB & 33700 & 83.37 & 84.27 & 86.70 & 85.48 & 84.23 & 33858 & 83.05 & 84.20 & 86.33 & 85.26 & 83.93 \\
        e\_All3\_resB & 33700 & 74.91 & 76.30 & 78.52 & 77.41 & 75.66 & 33858 & 75.44 & 77.63 & 78.58 & 78.11 & 76.21 \\
        \midrule
        
        e\_1st\_Dense & 33700 & 83.11 & 85.30 & 85.87 & 85.59 & 84.06 & 33858 & 84.00 & 85.50 & 87.01 & 86.26 & 84.91 \\
        e\_2nd\_Dense & 33700 & 83.72 & 86.44 & 86.18 & 86.31 & 84.71 & 33858 & 83.08 & 85.19 & 85.91 & 85.55 & 84.04 \\
        e\_bothDense & 33700 & 83.78 & 82.35 & 88.04 & 85.20 & 84.41 & 33858 & 83.66 & 83.46 & 87.45 & 85.46 & 84.42 \\
        \midrule
        
       \textbf{Whole arch.} & \textbf{33700} & \textbf{83.97} & \textbf{85.36} & \textbf{87.00} & \textbf{86.18} & \textbf{84.86} & \textbf{33858} & \textbf{84.53} & \textbf{84.85} & \textbf{87.98} & \textbf{86.41} & \textbf{85.33} \\
        \bottomrule
    \end{tabular}
\caption{Performance after removing different deep learning components of our architecture for model. e presents excluding the respective component of our architecture. Bold colored row presents the whole architecture's performance.}
\label{ablation_dl}
\end{table}

\subsection{Discussion}

Although the audio-only model with the meta-learner performed similarly to the multimodal model, the multimodal exploration provides valuable insight into the contribution of each sensing modality. The ablation analysis showed that audio is the dominant modality for this task, while additional smartwatch sensors provide limited performance gains. This is an important empirical finding, as it clarifies the relative contribution of each sensing modality. Furthermore, the results indicate that a minimal audio-only configuration can be sufficient for accurate interaction detection in this setting.

One plausible explanation for the dominance of audio is that our interaction definition is centered on vocal engagement, making audio the most directly relevant signal. Another contributing factor is the difference in temporal resolution: audio was sampled at 16{,}000~Hz, whereas the highest sampling rate among the other modalities was 25~Hz for PPG. As a result, within a 15-second window, audio provides substantially richer information, while modalities such as PPG may have insufficient samples to capture meaningful patterns. In addition, because the audio-only model already achieved high performance, there is limited room for additional modalities to further improve results. However, performance from individual sensors does not imply that non-audio modalities are uninformative. As shown in Table~\ref{multi_modal_1_modal_perf}, these modalities still capture some interaction-related signal, such as the 8.07\% sensitivity achieved by PPG. Rather, the results suggest that their contribution is substantially weaker than audio under the current 15-second window-level formulation.

The multimodal exploration also led to an architecture that enables interaction inference from a single 15-second sensing window. In contrast, the deployed audio-based framework relies on foreground speech and interaction cues accumulated over a candidate interaction interval, such that interactions can only be confirmed after sufficient evidence has been collected, typically after the interaction has ended. The window-level architecture (Figure~\ref{fig:multi_modal_arch}) therefore supports earlier interaction inference without requiring aggregation over the full interaction interval.

The architecture is also parameter-efficient. Excluding the meta-learner, the multimodal deep learning model contains approximately 0.63 million parameters and achieves 86.18\% balanced accuracy (Table~\ref{multi_modal_1_modal_perf}), while the audio-only model achieves similar performance using only 0.315 million parameters. In comparison, the deployed framework relies on YAMNet for both foreground speech detection and interaction-cue detection, where YAMNet alone has approximately 3.7 million parameters. To avoid confusion, we note that the final meta-learner configurations still use YAMNet-derived features and therefore have a larger effective parameter footprint. Thus, the parameter-efficiency claim refers specifically to the base deep learning architecture prior to the meta-learner.

Finally, the architecture, excluding the meta-learner, reduces dependence on explicit foreground speech extraction. Prior smartwatch-based social interaction detection systems often rely on dedicated foreground speech detection or feature extraction modules~\cite{Liang2023-gi,Zhang_Bertley_Liang_Thomaz_2025}. In contrast, our architecture (Figure~\ref{fig:multi_modal_arch}) learns directly from raw sensing streams without a separate foreground-speech-specific pipeline, while still achieving strong performance. This suggests that the model can learn interaction-relevant acoustic representations directly from short sensing windows, simplifying the inference pipeline and potentially reducing implementation complexity in future wearable deployments.

Detecting interactions from 15-second windows enables faster identification of interaction periods, including approximate start and end times. For example, consecutive windows classified as interactions can be used to estimate interaction boundaries. Although the 15-second window model has not yet been deployed on-device, its architecture does not introduce components that require substantial changes beyond the current deployment framework. For example, the audio-only model achieving 90.39\% balanced accuracy uses the same deep learning and machine learning libraries (e.g., Keras, TensorFlow, Scikit-learn) as the current on-watch system (Section~\ref{on_watch_imple_section}).

\section{Opportunities and Future Work}

Accurate recognition of social interactions in naturalistic settings enables applications across mental health, physical health and wellness, and human-centered computing. By demonstrating that interactions can be detected using lightweight, on-device sensing without storing raw audio, this work lowers barriers to deploying interaction-aware systems in real-world settings where privacy, battery life, and user burden are critical constraints.

\textit{Clinical Applications in Mental Health.} 
Our system provides a scalable tool for researchers studying social interaction patterns and their associations with mental and physical health outcomes (e.g., \cite{Leigh-Hunt_Bagguley_Bash_Turner_Turnbull_Valtorta_Caan_2017, ono2011relationship, pachucki2015mental}). With a balanced accuracy of 90.36\%, the system captures the majority of users’ interactions, enabling longitudinal measurement at a temporal resolution that is difficult to achieve using retrospective self-report alone. This capability supports applications such as monitoring social engagement and providing feedback to facilitate reflection or behavior change \cite{rabbi2015mybehavior}.

The system’s duty-cycled design further enables timely, context-aware interventions. Interaction-aware triggers could be used to prompt skill use or deliver support during or immediately following social interactions. By incorporating interaction cues beyond foreground speech and not requiring speech in every sensing window, the system may be particularly well suited for populations, such as individuals with social anxiety, for whom interactions may involve limited speech or extended silence, especially during early stages of exposure-based interventions \cite{Kampmann_Emmelkamp_Hartanto_Brinkman_Zijlstra_Morina_2016, de_Mooij_Fekkes_Miers_van_den_2023, info_doi_10_2196_13869}.

\textit{User Experience Applications.} 
Beyond clinical and research use, our findings have implications for smartwatch user experience design. Lightweight, on-device interaction detection enables notification and survey scheduling that respects social context without continuous sensing. For example, systems can defer interruptions until an interaction has concluded, reducing disruption and improving long-term usability and engagement.

\textit{Future Directions.}
Future work includes incorporating adaptive, user-specific thresholds; exploring richer yet privacy-preserving representations of social context; and validating the system across more diverse populations and environments. Integrating interaction detection with complementary self-report or contextual signals may further advance assessment of when interactions occur and also how they are experienced.

\section{Limitations}
This work has several limitations that reflect deliberate design tradeoffs, as well as opportunities for future research. First, our system does not store or transmit raw audio. All processing is performed on-device, and only derived features (e.g., log-mel spectrograms, foreground speech estimates, and interaction cues) are retained. While this design substantially reduces privacy risk, it limits the range of social interactions and analyses that can be supported. Second, the system is designed to detect interactions that are at least partially audio-based and does not capture text-based social interactions (e.g., texting or messaging on social media) or non-audio communication modalities (e.g., sign language).

Also, some noise in the labels likely remains, even though multiple findings support the reliability of the labels (findings suggesting the labels are reliable include that 60.14\% of notifications were answered within 1 minute, participant responses confirmed 584.35 cumulative hours with no missed interactions, analyses of participant-added interactions showed interaction cues at the beginning of reported intervals, and the system performed well in both pre-deployment and in-the-wild deployment). In addition, we used EMA-based labeling, a widely adopted approach for obtaining ground-truth data in real-world sensing studies. Nevertheless, labels may still be affected by factors such as recall bias, delayed responses, or participant-edited and added interactions. These sources of noise reflect practical tradeoffs when collecting annotations in real-world settings without imposing higher participant burden.

The system also relies on duty-cycled rather than continuous audio sensing. Although the 1.5-minute duty cycle balances privacy, battery consumption, and detection performance, it may miss brief interactions or interactions occurring entirely between probe windows. The current system could benefit from personalized adaptations to individual differences in speaking style, interaction patterns, or device placement in future work.

Finally, the evaluation was conducted primarily with participants from a single large public university. Although the sample included variation in age, roles, and daily routines, social interaction patterns and smartwatch usage may differ in other demographic, occupational, or cultural contexts. Further studies are needed to assess generalizability across more diverse populations and settings.

\section{Conclusion}
We presented the first privacy-preserving on-watch framework for detecting real-world social interactions using off-the-shelf smartwatches. The system leverages duty-cycled acoustic sensing to accurately identify interactions without storing or transmitting raw audio. To validate this approach, we conducted a real-world deployment with 38 participants wearing the smartwatch during their daily activities, yielding over 900 hours of data and thousands of annotated interaction events. Through extensive evaluation, we demonstrated that the proposed architecture, which predicts interactions using only 15 seconds of data, substantially improves interaction detection performance, achieving balanced accuracy exceeding 90\% while remaining computationally efficient and suitable for on-watch inference. By showing that reliable social interaction detection is possible under strict privacy, resource, and user-burden constraints, this work advances the design of scalable, social interaction-aware systems and opens new opportunities for applications in mental health, human-centered computing, and beyond.

\section{Dataset and Code Availability}

The interaction and non-interaction labels, together with the log-mel spectrogram audio features extracted during each duty cycle by the SocialPulse system, are publicly available at \href{https://zenodo.org/records/19767766}{Zenodo}. The complete preprocessing, analysis, and model development code is available at \href{https://zenodo.org/records/19973673}{Zenodo}.

\section*{Acknowledgments}

We thank Arafat Rahman for his assistance with testing early versions of the SocialPulse system.

This work was supported in part by a 3Cavaliers Seed Grant, a Thriving Youth in a Digital Environment (TYDE) Seed Grant, the National Institute of Mental Health of the National Institutes of Health under Award Number R01MH132138, and the Commonwealth Cyber Initiative (CCI), an investment in the advancement of cyber research, innovation, and workforce development. For more information about CCI, visit www.cyberinitiative.org.

\bibliographystyle{ACM-Reference-Format}
\bibliography{_reference}
\end{document}